\documentclass[11pt]{article}

\usepackage[top=60pt,bottom=70pt,left=75pt,right=75pt]{geometry}

\usepackage{amsmath, amssymb, amsfonts, latexsym, euscript, mathalfa, mathrsfs}
\usepackage{graphicx, color, epsfig, xcolor}
\usepackage{cite}
\usepackage{empheq}

\usepackage{setspace, sectsty}

\usepackage{braket, cancel}

\usepackage[debug,pageanchor=false]{hyperref}
\definecolor{link}{rgb}{.8,.15,.1}
\hypersetup{colorlinks=true,linkcolor=link,citecolor=link,urlcolor=link,linktocpage}

\makeatletter
\@addtoreset{equation}{section}
\makeatother

\newlength{\sswidth}

\newcommand{\nn}{\nonumber}

\newcommand{\R}{\mathrm{Re}}
\newcommand{\I}{\mathrm{Im}}
\newcommand{\ncal}{\mathcal{N}}
\newcommand{\vol}{\mathrm{vol}}

\newcommand{\g}{\gamma}

\newcommand{\fpm}{\sqrt{f_+f_-}}
\newcommand{\KE}{{\rm KE}_4}
\newcommand{\Of}{O4-plane singularity }

\newcommand{\CC}{\mathcal{C}}
\newcommand{\s}{\sigma}

\renewcommand{\a}{\alpha}

\newcommand{\eq}[1]{\begin{equation}\begin{split}#1\end{split}\end{equation}}
\newcommand{\sub}[1]{\begin{subequations}\begin{align}#1\end{align}\end{subequations}}

\begin{document}

\begin{titlepage}

\begin{flushright} \small
UUITP-20/18
\end{flushright}

\begin{center}

\noindent

{\Large \bf A massive class of $\mathcal{N} = 2$ AdS$_4$ IIA solutions}

\bigskip\medskip

Achilleas Passias$^1$, Dani\"{e}l Prins$^{2,3}$ and Alessandro Tomasiello$^3$\\

\bigskip\medskip
{\small

$^1$Department of Physics and Astronomy, Uppsala University,\\
Box 516, SE-75120 Uppsala, Sweden
\\	
\vspace{.3cm}
$^2$Institut de Physique Th\'{e}orique, Universit\'{e} Paris Saclay, CNRS, CEA, \\
F-91191 Gif-sur-Yvette, France
\\	
\vspace{.3cm}
$^3$Dipartimento di Fisica, Universit\`a di Milano--Bicocca, \\ Piazza della Scienza 3, I-20126 Milano, Italy \\ and \\ INFN, sezione di Milano--Bicocca
}

\vskip .5cm 
{\small \tt achilleas.passias@physics.uu.se, daniel.prins@cea.fr, alessandro.tomasiello@unimib.it}

\vskip .9cm 
     	{\bf Abstract }
\vskip .1in
\end{center}
We initiate a classification of $\ncal =2$  supersymmetric AdS$_4$ solutions of (massive) type IIA supergravity.
The internal space is locally equipped with either an SU(2) or an identity structure.
We focus on the SU(2) structure and determine the conditions it satisfies, dictated by supersymmetry.
Imposing as an Ansatz that the internal space is complex,
we reduce the problem of finding solutions to a Riccati ODE, which we solve analytically.
We obtain in this fashion a large number of new families of solutions, both regular as well as with localized O8-planes and conical Calabi--Yau singularities.
We also recover many solutions already discussed in the literature.

\noindent

\noindent

\vfill
\eject

\end{titlepage}

\tableofcontents

\section{Introduction}

The study of four-dimensional anti-de Sitter solutions of string/M-theory is of considerable interest both in the context of flux compactifications and of the AdS/CFT correspondence. The prototypical class of such solutions is the Freund--Rubin class \cite{freund-rubin} in M-theory, where the flux is along the anti-de Sitter spacetime and the internal manifold is Einstein. This class allows for various amounts of supersymmetry, which
further constrain the geometry of the internal manifold $M_7$.  Maximal supersymmetry imposes $M_7 \simeq S^7$, while
with $\mathcal{N}=2$ supersymmetry, which is the focus of this paper, $M_7$ is Sasaki--Einstein. Outside of this class very few solutions are known in M-theory \cite{Corrado:2001nv, gabella-martelli-passias-sparks, halmagyi-pilch-warner}.

From a holographic perspective, AdS$_4$ solutions are dual to Chern--Simons--matter field theories in three dimensions. A
good control of the correspondence typically requires extended supersymmetry, and $\mathcal{N}=2$ provides a nice
balance between control and variety. When in M-theory (and its type IIA reduction), the sum of the Chern--Simons levels of the gauge groups that characterize the field theory is zero. A non-zero sum corresponds to a non-zero Romans mass  in type IIA string theory \cite{gaiotto-t,fujita-li-ryu-takayanagi}. So far, almost all known $\mathcal{N}=2$ solutions of
massive type IIA supergravity are numerical  \cite{petrini-zaffaroni,lust-tsimpis-singlet-2,ajtz,tz}, with one notable exception being the Guarino--Jafferis--Varela solution \cite{guarino-jafferis-varela}. In this paper, we overturn this status,
finding a vast number of analytic solutions.

We will analyze the constraints imposed by supersymmetry by employing the ``pure spinor'' method, originally devised
for $\mathcal{N}=1$ solutions \cite{gmpt2,gmpt3}. Adapting this method to ${\cal N}=2$ is not entirely straightforward. Until recently, in most of the literature one has resorted to imposing, on top of ${\cal N}=1$ supersymmetry, the presence of a vector field that leaves all fields but the supersymmetry parameters  invariant, thus representing the R-symmetry action. This has proven useful, but has to be supplemented by an inspired Ansatz, and so far has resulted in the aforementioned numerical solutions.

A different approach has been put forward in \cite{passias-solard-t}, based on the work of \cite{10d}. In the latter reference, the conditions for supersymmetry were expressed in terms of differential forms in the spirit of generalized geometry, without further assumptions on the form of the solution. In \cite{passias-solard-t}, these were adapted to the specific case of $\ncal =2$ AdS$_4 \times M_6$ solutions of type IIB supergravity, obtaining a set of ${\cal N}=2$ pure spinor equations. After some work this resulted in a system of partial differential equations which characterize all possible solutions. In this paper we apply the same idea to type IIA supergravity.

The set of ${\cal N}=2$ pure spinor equations we obtain superficially resembles that of \cite{passias-solard-t} in IIB; as is the case for ${\cal N}=1$, it is obtained by exchanging odd with even pure spinors. However, the geometric constraints that follow differ early on in the analysis. While in IIB the structure group on $M_6$ is exclusively the identity, in IIA it can be either the identity or SU(2).\footnote{This parallels what happens for ${\cal N}=1$: IIA solutions can have SU(3) or SU(2) structure, while in IIB only the latter is allowed \cite{Behrndt:2005bv}.} In this paper we will focus on the SU(2) structure case, leaving the identity structure for future work.\footnote{$\mathcal{N}=2$ solutions with SU(2) structure have been previously studied in \cite{Bovy:2005qq}.}
The set of constraints we obtain from supersymmetry on the SU(2) structure also imply the Bianchi identities for the form fields, and all equations of motion.

Within the SU(2) structure case, we find two classes which we call ``class K'' and ``class HK'', because the internal manifold $M_6$ contains a four-dimensional subspace $M_4$ equipped with either a K\"ahler or a hyper-K\"ahler metric. We work out the supersymmetry constraints in full detail for both classes. Class HK leads to one local metric. On the other hand, class K  leads to a very rich structure of solutions.

In particular, a simple and natural Ansatz (inspired by \cite{gauntlett-martelli-sparks-waldram-M}) is that $M_6$ admits a complex structure. After imposing this, the problem of finding solutions reduces to a single  ordinary differential equation (ODE) of Riccati type. The analysis is further subdivided according to whether $M_4$ is K\"ahler--Einstein, or is a product of two Riemann surfaces. In both cases we find the most general solution to the ODE analytically.

This results in two new families of analytic solutions. The fully regular ones are identified with the numerical solutions we mentioned above \cite{petrini-zaffaroni,lust-tsimpis-singlet-2,ajtz,tz}. In the massless limit, they result in the IIA reduction of various previously-known Sasaki--Einstein manifolds: the so-called $Y^{p,k}$ \cite{gauntlett-martelli-sparks-waldram-SE7}, a generalization thereof called $A^{p,q,r}$ \cite{gauntlett-martelli-sparks-waldram-Apqr,chen-lu-pope-vazquezporitz,tz}, and the older $M^{3,2}$ \cite{witten-M32,castellani-dauria-fre-M32}, $Q^{1,1,1}$ \cite{dauria-fre-vannieuwenhuizen-Q111}, as well as the Fubini--Study metric on $\mathbb{CP}^3$. As already suggested by numerical evidence in \cite{ajtz}, a certain ``maximal'' massive deformation exists, which results (when $M_4= \mathbb{CP}^2$) in the Guarino--Jafferis--Varela solution \cite{guarino-jafferis-varela}.

Beyond this maximal deformation, the solutions still exist, but develop singularities that have a physical interpretation as corresponding to the presence of various orientifold planes. In most cases these are smeared in some directions and localized in others; some limits of the parameters however produce solutions with fully localized O8-planes.

The rest of the paper is structured as follows. In section \ref{sec:reduction} we specialize the system of ten-dimensional equations obtained in \cite{10d} to ${\cal N}=2$ AdS$_4$ IIA solutions. As we mentioned, the analysis is similar to the one in IIB \cite[Sec.~2,3]{passias-solard-t}, but some differences begin to emerge already here, and in particular we see that the SU(2) structure case is admissible, which we then focus on.
In section \ref{sec:analysis} we analyze the system for this case by eliminating redundancies and obtaining the geometrical consequences of the system; the two classes K and HK are analyzed in turn. Being class HK rather limited, we devote section \ref{cohom1} to class K, under the assumption that $M_6$ admits a complex structure. As anticipated, we obtain two main families of analytic solutions, depending on several parameters. We summarize our findings in section \ref{sub:sum}.

\section{Reduction of the ten-dimensional supersymmetry equations}
\label{sec:reduction}

In this section we will specialize the system of equations obtained in \cite{10d} as a set of necessary and sufficient conditions for any ten-dimensional solution of type II supergravity to preserve superymmetry, to the case of an AdS$_4$ background of type IIA supergravity preserving $\mathcal{N}=2$ supersymmetry. The process is similar to the one followed for type IIB supergravity in
\cite{passias-solard-t} and we refer the reader there for more details, especially on conventions.

We begin by reviewing the system of equations of \cite{10d}, which are summarized in section 3.1 of that paper, focusing on the following subset of equations:
\begin{subequations}\label{eq:10deq}
\begin{align}
&d_H(e^{-\phi} \Phi) = -(\widetilde K \wedge + \iota_K) F_{(10d)} \ , \label{10d-eq1} \\
&d \widetilde K = \iota_K H \, ,\qquad \nabla_{(M} K_{N)} = 0 \ . \label{10d-eq2}
\end{align}
\end{subequations}
Here $\phi$ is the dilaton, $H$ is the NS--NS three-form field strength, $d_H \equiv d - H\wedge$, and $F_{(10d)}$ is an even form obtained as a formal sum of all the R--R field strengths.  $\lambda(F_p) \equiv (-1)^{\lfloor p/2\rfloor} F_p$, for $F_p$ a $p$-form.

Moreover, $\Phi$ in (\ref{eq:10deq}) is an even form that corresponds via the Clifford map $\gamma^{M_1\ldots M_k}\mapsto dx^{M_1}\wedge \ldots \wedge dx^{M_k}$ to the bispinor $\epsilon_1 \otimes \overline{\epsilon_2}$, where $\epsilon_{1,2}$ are the parameters of the supersymmetry transformations (which we take to be two Majorana--Weyl spinors of positive and negative chirality respectively). The vector $K$ and the one-form $\widetilde K$ are defined by
\begin{equation}
K \equiv \tfrac{1}{64}(\overline{\epsilon_1} \Gamma^M \epsilon_1 + \overline{\epsilon_2} \Gamma^M \epsilon_2) \partial_M \, , \qquad
\widetilde K \equiv \tfrac{1}{64}(\overline{\epsilon_1} \Gamma_M \epsilon_1 - \overline{\epsilon_2} \Gamma_M \epsilon_2) dx^M \, .
\end{equation}
As we see in the second equation of (\ref{10d-eq2}), $K$ is a Killing vector and is in fact a symmetry of all fields in a solution.

We will apply (\ref{eq:10deq}) to AdS$_4$ solutions, meaning that we will take all fields to preserve its SO$(3, 2)$ isometry group. In particular we will take the metric to be of the warped product form:
\begin{equation}
ds^2_{10} = e^{2A} ds^2_{\mathrm{AdS}_4} + ds^2_{M_6} \,.
\end{equation}
The symmetry requirement also implies that the $H$ field will only be a form on $M_6$, while the R--R field strengths will be decomposed as
\begin{equation}
F_{(10d)} = e^{4A} \vol_4 \wedge \star_6 \lambda(F) + F \ ,
\qquad F = F_0 + F_2 + F_4 + F_6 \ .
\end{equation}

The supersymmetry parameters $\epsilon_{1,2}$ will also be sums of tensor products of spinors on AdS$_4$ and $M_6$.
For $\mathcal{N}=2$ supersymmetry, the decomposition reads
\begin{subequations}\label{spin_decomp}
\begin{align}
\epsilon_1 &= \sum_{I = 1}^2 \chi_+^I \otimes \eta_{1+}^I + \sum_{J = 1}^2 \chi_-^J \otimes \eta_{1-}^J
\ , \\
\epsilon_2 &= \sum_{I = 1}^2 \chi_+^I \otimes \eta_{2-}^I + \sum_{J = 1}^2 \chi_-^J \otimes \eta_{2+}^J
\ .
\end{align}
\end{subequations}
The $\chi$'s are a basis of AdS$_4$ Killing spinors:
\begin{equation}\label{Killing}
\nabla_\mu \chi^I_\pm = \frac{1}{2} \gamma_\mu \chi^I_\mp \ ,
 \qquad \nabla_\mu \overline{\chi^I_\pm} = - \frac{1}{2} \overline{\chi^I_\mp} \gamma_\mu \ ,
\end{equation}
where $\mu=0,\ldots,3$. We will assume $\chi^1_+$ and $\chi^2_+$ to be linearly independent, since otherwise we would only have $\mathcal{N} = 1$ supersymmetry. The gamma matrices decompose according to
\begin{equation}\label{cliff_decomp}
\Gamma_\mu = e^A \gamma_\mu \otimes \mathbb{I} \ , \qquad
\Gamma_{m + 3} = \g_5 \otimes \gamma_m \ , \qquad
\Gamma_{11} \equiv \Gamma^0 \dots \Gamma^{9} = \gamma_5 \otimes \gamma_7 \ ,
\end{equation}
with $m=1,2,\dots 6$. $\g_5$ and $\gamma_7$ are the external and internal chirality operators.

Let us now reduce (\ref{eq:10deq}) for this case of ${\cal N}=2$ AdS$_4$ solutions.
We start with \eqref{10d-eq2}: $\widetilde{K}$, $K$ decompose as
\begin{subequations}
\begin{align}
\widetilde{K} _\mu &= \frac{e^A}{32}\sum_{I,J=1}^{2} \overline{\chi^I_+} \gamma_\mu \chi^J_+  (\overline{\eta^I_{1+}} \eta^J_{1+} - \overline{\eta^I_{2-}} \eta^J_{2-})\, ,&
\widetilde{K}_m &= -\frac{1}{16} \R (\overline{\chi^1_+} \chi^2_- \tilde\xi_m) \ , \\
K^\mu &= \frac{e^{-A}}{32}\sum_{I,J=1}^{2} \overline{\chi^I_+} \gamma^\mu \chi^J_+ (\overline{\eta^I_{1+}} \eta^J_{1+} + \overline{\eta^I_{2-}} \eta^J_{2-})\, ,&
K^m &= -\frac{1}{16} \R (\overline{\chi^1_+} \chi^2_- \xi^m) \ ,
\end{align}
\end{subequations}
where
\begin{equation}
\tilde\xi_m \equiv \overline{\eta^1_{1+}} \gamma_m \eta^2_{1-} - \overline{\eta^1_{2-}} \gamma_m \eta^2_{2+} \ , \qquad \xi^m \equiv \overline{\eta^1_{1+}} \gamma^m \eta^2_{1-} + \overline{\eta^1_{2-}} \gamma^m \eta^2_{2+} \ .
\end{equation}
We thus find that (\ref{10d-eq2}) gives
\begin{equation}\label{spinor_scalars}
\overline{\eta_{1+}^{(I}} \eta_{1+}^{J)} =
\overline{\eta_{2+}^{(I}} \eta_{2+}^{J)} \equiv \frac{1}{2} c^{IJ} e^A \ , \qquad
-i\overline{\eta_{1+}^{[I}} \eta_{1+}^{J]} = -i\overline{\eta_{2+}^{[I}} \eta_{2+}^{J]} \equiv \epsilon^{IJ} f \ ,
\end{equation}
where $c^{IJ}$ are constants, and
\begin{equation}
d(e^A f)  =  -\frac{1}{2} \I \tilde \xi  \,, \qquad
d \tilde \xi = i_\xi H \, ,\qquad
\I (\xi) = 0 \,, \qquad \nabla_{(n} \xi_{m)} = 0 \,.
\end{equation}
Hence $\xi$ is a Killing vector and in fact realizes the R-symmetry, acting on the $I$ index in (\ref{spin_decomp}). Notice that there are subtle sign differences in these formulas with respect to similar ones in IIB \cite{passias-solard-t}.

To reduce \eqref{10d-eq1}, we write $\Phi$ as a wedge product of external and internal forms:
\begin{equation}
\Phi = \sum_{IJ} (
- \chi^I_+\overline{\chi^J_+} \wedge \eta^I_{1+} \overline{\eta^J_{2-}} +
  \chi^I_+\overline{\chi^J_-} \wedge \eta^I_{1+} \overline{\eta^J_{2+}} +
  \chi^I_-\overline{\chi^J_+} \wedge \eta^I_{1-} \overline{\eta^J_{2-}} +
  \chi^I_-\overline{\chi^J_-} \wedge \eta^I_{1-} \overline{\eta^J_{2+}}
) \ .
\end{equation}
Again we use bispinors to denote the corresponding polyform under the Clifford map.  Using  \eqref{Killing} we can derive the exterior differential of the four-dimensional spacetime polyforms:
\begin{subequations}
\begin{align}
d(\chi_{\pm}^{I}\overline{\chi_{\pm}^{J}})
&= 2 \sum_k\left(1-\tfrac{1}{4}(-1)^k(4-2k)\right) \R (\chi_{\mp}^{I}\overline{\chi_{\pm}^{J}})_k \ , \\
d(\chi_{\pm}^{I}\overline{\chi_{\mp}^{J}})
&= 2i\sum_k\left(1+\tfrac{1}{4}(-1)^k(4-2k)\right) \I (\chi_{\mp}^{I}\overline{\chi_{\mp}^{J}})_k \ ,
\end{align}
\end{subequations}
where $k$ is the form degree. Using this in (\ref{10d-eq1}) and factoring spacetime forms, we get purely internal equations:
\begin{subequations}\label{eq:pureIJ}
\begin{align}
d_H\left(e^{2A-\phi} \phi^{(IJ)}_+\right) - 2 e^{A-\phi} \R \phi_-^{(IJ)} &= 0 \ , \label{eq:pureIJ1} \\
d_H\left(e^{3A - \phi} \R \phi_-^{[IJ]}\right) &= 0 \ , \label{eq:pureIJ2}\\
d_H\left(e^{A-\phi} \I \phi^{[IJ]}_-\right) - e^{-\phi} \I \phi_+^{[IJ]} &= \frac{1}{8} e^A f F \epsilon^{IJ}\ ;
\label{RRflux}
\end{align}
and
\begin{align}
d_H\left(e^{3A-\phi} \I \phi_-^{(IJ)}\right) - 3 e^{2A-\phi} \I \phi_+^{(IJ)} &= \frac{1}{16} c^{IJ} e^{4A} \star \lambda(F) \ , \label{eq:pureIJ4} \\
d_H\left(e^{-\phi} \phi_+^{[IJ]}\right) &= -\frac{1}{16}(\bar{\tilde \xi} \wedge + \iota_\xi) F \epsilon^{IJ} \ , \label{eq:pureIJ5} \\
d_H\left(e^{4A-\phi} \phi_+^{[IJ]}\right) - 4 e^{3A-\phi}  \R \phi_-^{[IJ]}&= -\frac{i}{16}(\bar{\tilde \xi} \wedge + \iota_\xi) e^{4A} \star \lambda(F)  \epsilon^{IJ}\ ,
\label{eq:pureIJ6}
\end{align}
\end{subequations}
where
\begin{equation}\label{eq:phiIJ}
\phi_+^{IJ} \equiv \eta^I_{1+}\overline{\eta^J_{2+}}
\ , \qquad
\phi_-^{IJ} \equiv \eta^I_{1+}\overline{\eta^J_{2-}} \ ,
\end{equation}
and $\bar{\tilde \xi}$ is the complex conjugate of $\tilde{\xi}$.

As was the case for the corresponding system of equations in type IIB supergravity \cite{passias-solard-t}, the system (\ref{eq:pureIJ}), although alarmingly large, has a high degree of redundancy. For instance, we will see soon that $c^{IJ}$ can be set proportional to the identity; after that one can see that the equations that involve the R--R fields are redundant,  except for \eqref{RRflux}. Also, the $I\neq J$ components are redundant, since the $I=J$ ones furnish two copies of the pure spinor equations \cite{gmpt2,gmpt3} for ${\cal N}=1$ AdS$_4$ solutions.
Finally, the remaining ``pairing equations'' \cite[(3.1c,d)]{10d} are redundant as for \cite{passias-solard-t}.

In spite of this redundancy, (\ref{eq:pureIJ}) will be more convenient for our analysis than a repeated application of the ${\cal N}=1$ equations \cite{gmpt3}.

\subsection{Parametrization of the pure spinors}
\label{sec:param}

In this section we will parametrize the pure spinors $\phi_\pm^{IJ}$ in terms of a set of differential forms.

Before introducing the parametrization, we will fix the constants $c^{IJ}$ of \eqref{spinor_scalars} as
\begin{equation}\label{eq:cIJd}
c^{IJ} = 2 \delta^{IJ} \ ,
\end{equation}
where $\delta^{IJ}$ is the Kronecker delta. This is permitted as the decomposition Ansatz \eqref{spin_decomp} sets the internal spinors only up to a GL$(2,\mathbb{R})$ transformation that leaves invariant the norms $\|\eta_{i+}^I\|$ (which are equal to $e^A$, by \eqref{spinor_scalars}). The details of this transformation can be found in \cite{passias-solard-t}.
Since $c^{12} = \overline{\eta_{i+}^{(1}} \eta_{i+}^{2)} = 0$, from $\overline{\eta_{i+}^{[1}} \eta_{i+}^{2]} = \overline{\eta_{i+}^{1}} \eta_{i+}^{2}$ and $|\overline{\eta^1_{i+}}\eta^2_{i+}| \leq \sqrt{\|\eta^1_{i+}\|\|\eta^2_{i+}\|}$ it follows that
\begin{equation}\label{f_ineq}
|f| \leq e^A \ .
\end{equation}

Furthermore, instead of $\eta^I_{i+}$ we will work with
\begin{equation}\label{eq:etapm12}
\eta^\pm_{i+} = \frac{1}{\sqrt{2}} (\eta^1_{i+} \pm i \eta^2_{i+})
\end{equation}
which have charge $\pm 1$ under the U$(1)\simeq$ SO$(2)$ R-symmetry.
From \eqref{spinor_scalars} and (\ref{eq:cIJd}) we then have
\begin{equation}\label{spinor_scalarsII}
\overline{\eta^\pm_{i+}} \eta^\mp_{i+} = 0 \ , \qquad
\overline{\eta^\pm_{i+}} \eta^\pm_{i+} = f_\mp\equiv e^A \mp f \ .
\end{equation}

The internal spinors $\eta_{i+}^\pm$ can be parametrized in terms of a chiral spinor $\eta_+$ of positive chirality (and its complex conjugate $\eta_- \equiv (\eta_+)^c)$ as follows:
\begin{subequations}\label{eq:etapm-param}
\begin{align}
\eta^+_{1+} &= \sqrt{f_-} \eta_+ \ ,&\eta^+_{2+} &= \sqrt{f_-} \left(a \eta_+ + \frac{1}{2} b w_3 \eta_-\right) \ , \\
\eta^-_{1+} &= \sqrt{f_+} \frac{1}{2} w_1 \eta_- \ ,&
\eta^-_{2+} &= \sqrt{f_+} \frac{1}{2} c w_2\left(a^* \eta_- - \frac{1}{2} b \overline{w_3}\eta_+\right) \ ,
\end{align}
\end{subequations}
where the $w_i$ are one-forms, $a$ is a function taking value in $ \mathbb{C}$ and $b, c$ are real functions. The latter satisfy
\begin{align}\label{abc}
|a|^2 + b^2 = 1 \,,  \qquad
c^{-1} = \left(|z_1|^2b^2+|a|^2\right)^{1/2} \,, \qquad z_1 \equiv \frac{1}{2} \overline{w_2} \cdot w_3\,,
\end{align}
where $\cdot$ denotes the inner product.

The chiral spinor $\eta_+$ defines an SU(3) structure, characterized by a real two-form $J$ and a holomorphic three-form $\Omega$, as
\begin{equation}
\eta_+\overline{\eta_+} = \frac{1}{8} e^{-iJ} \ , \qquad
\eta_+\overline{\eta_-} = -\frac{1}{8} \Omega \ ,
\end{equation}
with $J$, $\Omega$ satisfying $J \wedge \Omega = 0$ and $J \wedge J \wedge J = \frac{3}{4} i \Omega \wedge \overline{\Omega}$.

When they are not all linearly dependent, the one-forms $w_i$ parametrize an identity structure and are holomorphic with respect to the almost complex structure $J$ defined by $\eta_+$. We will leave this generic case to future work; in this paper, we will limit ourselves to analyzing the case of an SU$(2)$ structure, for which the $w_i$ are all linearly dependent. Such a case is not guaranteed to be compatible with the supersymmetry equations a priori, and indeed it is not allowed
in type IIB supergravity \cite{passias-solard-t}. However, as we will see, in type IIA supergravity solutions with $\mathcal{N}=2$ supersymmetry
and an SU$(2)$ structure do exist.

We will thus take
\begin{equation}
w_2 = z_3 w_1 \ , \qquad
w_3 = z_2^* w_1 \ , \qquad
 \{z_2\, ,z_3 \in C(M_6,\mathbb{C}) : |z_2| = |z_3| = 1 \}  \ ,
\end{equation}
and set $w_1 \equiv w$, with normalized norm $||w||^2 = 2$. Note that $z_1 \equiv \frac{1}{2} \overline{w_2} \cdot w_3 = z_2^* z_3^*$, hence
$|z_1|^2=1$ and $c=1$.
The SU$(2)$ structure is defined by the one-form $w$, a real two-form $j$ and a holomorphic two-form $\omega$, with
\begin{equation}
J = j + \frac{i}{2} w \wedge \overline{w} \ , \qquad \Omega = w \wedge \omega \ ,
\end{equation}
and $w$, $j$ and $\omega$ satisfying $\iota_w j = 0 = \iota_w \omega$, $j \wedge \omega = 0$ and $j \wedge j = \frac{1}{2} \omega \wedge \overline{\omega}$.

We can now express the pure spinors
\begin{equation}\label{eq:phipmpm}
\phi_+^{\pm\pm} \equiv \eta^\pm_{1+}\overline{\eta^\pm_{2+}}
\ , \qquad
\phi_-^{\pm\pm} \equiv \eta^\pm_{1+}\overline{\eta^\pm_{2-}} \ ,
\end{equation}
in terms of forms:
\begin{subequations}
\begin{align}
\phi^{++}_+ &= \frac{1}{8}f_-\left[a^* \left( e^{-ij} + \frac{1}{2} w \wedge \overline{w}\wedge e^{-ij} \right) + \frac{1}{2} b z_2 (\overline{w} \wedge w \wedge \omega - 2 \omega)\right] \ , \\
\phi^{++}_- &=  \frac{1}{8}f_-\left[- a w \wedge \omega - b z_2^* w \wedge e^{-ij} \right] \ , \\
\phi^{+-}_+ &= \frac{1}{8}\fpm \left[\frac{1}{2} a z_3^* (\overline{w} \wedge w \wedge \omega - 2\omega) - b  z_3^* z_2^* \left( e^{-ij} + \frac{1}{2} w \wedge \overline{w}\wedge e^{-ij} \right) \right] \ , \\
\phi^{+-}_- &= \frac{1}{8}\fpm\left[-a^*  z_3 w \wedge e^{-ij} + b z_3 z_2 w \wedge \omega\right] \ , \\
\phi^{-+}_+ &= \frac{1}{8}\fpm \left[\frac{1}{2}a^*(w \wedge \overline{w} \wedge \overline{\omega}+2\overline{\omega}) + b z_2 \left( e^{ij} + \frac{1}{2} w \wedge \overline{w}\wedge e^{ij} \right) \right] \ , \\
\phi^{-+}_- &= \frac{1}{8} \fpm \left[a w \wedge e^{ij} - b z_2^* w \wedge \overline{\omega})\right] \ , \\
\phi^{--}_+ &= \frac{1}{8} f_+ \left[a  z_3^* \left( e^{ij} + \frac{1}{2} w \wedge \overline{w} \wedge e^{ij} \right) - \frac{1}{2} b z_3^* z_2^* (w \wedge \overline{w} \wedge \overline{\omega}+2\overline{\omega})\right] \ , \\
\phi^{--}_- &= \frac{1}{8}f_+ \left[- a^* z_3 w \wedge \overline{\omega} - b z_3 z_2 w \wedge e^{ij}\right]
\end{align}
\end{subequations}
We also have
\begin{subequations}
\begin{align}
(\xi)^\flat &= i\sqrt{f_+f_-}\left(\overline{w} + z_3 w \right) \ , \label{xidual} \\
\tilde{\xi} &= i\sqrt{f_+f_-}\left(\overline{w} - z_3 w \right) \ ,
\end{align}
\end{subequations}
where $(\xi)^\flat$ is the one-form dual to the vector $\xi$.

\subsection{System of equations}
In terms of the pure spinors $\phi_\pm^{\pm\pm}$ introduced in  (\ref{eq:phipmpm}),
the system of supersymmetry equations (\ref{eq:pureIJ}) reads
\begin{subequations}\label{SUSYeq}
\begin{align}
d_H\left[e^{2A-\phi}\phi_+^{+-}\right] -  e^{A-\phi} (\phi_-^{++} + \overline{\phi_-^{--}}) &=0  \ ,
\label{S1} \\
d_H\left[e^{2A-\phi}(\phi_+^{++} + \phi_+^{--})\right] - 2 e^{A-\phi} \R(\phi_-^{+-} + \phi_-^{-+}) &=0  \ ,
\label{S2} \\
d_H\left[e^{2A-\phi}\phi_+^{-+}\right] -  e^{A-\phi} (\overline{\phi_-^{++}} + \phi_-^{--}) &=0  \ ,
\label{S3} \\
d_H\left[e^{3A-\phi}\I(\phi_-^{+-} - \phi_-^{-+})\right] &=0 \ ,
\label{S4} \\
d_H\left[e^{3A-\phi}(\phi_-^{++} - \overline{\phi_-^{--}})\right] - 3 e^{2A-\phi}(\phi_+^{+-} - \overline{\phi_+^{-+}}) &=0  \ ,
\label{S5} \\
d_H\left[e^{A-\phi} \R(\phi_-^{+-} - \phi_-^{-+})\right] - e^{-\phi} \R(\phi_+^{++} - \phi_+^{--}) &= \frac{1}{4} e^A f F  \ ,
\label{F2}
\end{align}
\end{subequations}
and
\begin{subequations}\label{extraF}
\begin{align}
d_H\left[e^{3A-\phi} \I(\phi_-^{+-} + \phi_-^{-+})\right] - 3 e^{2A-\phi} \I (\phi_+^{++} + \phi_+^{--}) &= \frac{1}{4} e^{4A} \star \lambda(F) \ , \label{F1}  \\
d_H\left[e^{-\phi}(\phi_+^{++} - \phi_+^{--})\right] &= \frac{i}{8}(\bar{\tilde \xi} \wedge + \iota_\xi) F  \ ,
\label{F3} \\
d_H\left[e^{4A-\phi}(\phi_+^{++} - \phi_+^{--})\right] -
4 i e^{3A-\phi} \I(\phi_-^{+-} - \phi_-^{-+}) &= -\frac{1}{8}(\bar{\tilde \xi} \wedge + \iota_\xi) e^{4A} \star \lambda(F)  \ .
\label{F4}
\end{align}
\end{subequations}
We also have
\begin{subequations}
\begin{align}
\I (\xi) &= 0 \ , \label{X1} \\
d(e^A f) + \frac{1}{2} \I(\tilde \xi)  &= 0 \ , \label{X2} \\
d \tilde \xi - i_\xi H &= 0 \ , \label{X3} \\
\nabla_{(n} \xi_{m)} &= 0 \label{X4} \ ,
\end{align}
\end{subequations}
which were obtained from \eqref{10d-eq2} and the condition that the ten-dimensional vector $K$ is Killing.

We are showing \eqref{extraF} separately because they are in fact implied by \eqref{SUSYeq}. Even if they are redundant, \eqref{F1} and \eqref{F3} are useful to show that the equations of motion and the Bianchi identities of the R--R fields are automatically satisfied; see also our comment after (\ref{eq:phiIJ}).

Acting with $d_H$ on \eqref{F1}, and using the imaginary part of \eqref{S2} it follows that
\begin{equation}
d_H(e^{4A} \star\lambda(F)) = 0 \ ,
\end{equation}
which are the equations of motion. Acting with $d_H$ on \eqref{F2}, using \eqref{X2}, and subtracting the real part of \eqref{F3}, it follows that
\begin{equation}
d_H F = 0 \ ,
\end{equation}
which are the Bianchi identities of the R--R fields. This holds under the assumption that the Bianchi identity for $H$, $dH=0$, is satisfied. Although it is not immediately obvious, we shall see that the NS--NS Bianchi identity is in fact implied by the supersymmetry equations.

\section{Analysis of the supersymmetry equations}
\label{sec:analysis}

In this section we analyze the supersymmetry equations obtained in section \ref{sec:reduction}.
As we anticipated, not all the equations are independent, and we will be able to reduce them to a significantly smaller set which characterizes
the SU$(2)$ structure on the internal manifold.

We will distinguish two cases. This is because certain equations, such as the zero-form component of (\ref{S3}), have an overall factor of $b$, and can thus be solved either by setting $b=0$ or by keeping $b\neq0$ and setting to zero the remaining factor. It turns out that these two cases are qualitatively different, and we will consider them in separate subsections.

We will refer to the first case as ``Class K'' and the second one
as ``Class HK'', because in these two cases $M_6$ will turn out to contain respectively a K\"ahler and a hyper-K\"ahler four-dimensional submanifold.

\subsection{Class K}
\label{sub:K}
In this section we look at the case $b = 0$. The condition \eqref{X1}, $\I(\xi) = 0$, fixes $z_3 = -1$, while \eqref{X2} gives
\begin{equation}
d(e^Af) = - \sqrt{f_+f_-} \R w \ .
\end{equation}
We define $y \equiv e^A f$, which we will use as a coordinate, so that
\begin{equation}
\R w = -\frac{1}{\sqrt{f_+f_-}} dy \ ,
\end{equation}
where now $f_+f_- = e^{2A} - f^2 = e^{2A} - e^{-2A} y^2$. We will also introduce a coordinate $\psi$, adapted to the Killing vector as
\begin{equation}
\xi = 4 \partial_\psi \ .
\end{equation}
From \eqref{xidual} it follows that
\begin{equation}
\I w = \frac{1}{2} \sqrt{f_+f_-} (d\psi + \rho) \ ,
\end{equation}
where $\rho$ is a one-form on the four-dimensional subspace orthogonal to $w$.

The zero-form component of \eqref{F2} yields
\begin{equation}
\R a = - e^{-A+\phi} y F_0 \ ,
\end{equation}
while the one-form part of \eqref{S1}--\eqref{S5} give
\begin{equation}
d(e^{3A-\phi} \I a) = 0 \ .
\end{equation}
We can thus write
\begin{equation}\label{Aa}
a = e^{-A+\phi}(- y F_0 + i e^{-2A} \ell) \ , \qquad \ell = {\rm constant} \ .
\end{equation}
Note that from the above expression and \eqref{abc}, which for $b=0$ yields $|a|^2=1$, it follows that $F_0$ and $\ell$ cannot be
simultaneously zero.

Given the above we find that the two-form part of \eqref{S1}--\eqref{S5} is automatically satisfied, while the three-form part yields:
\begin{subequations}
\begin{align}
F_0 d(e^{2A} y j - y^2 \R w \wedge \I w) + \ell H &=0 \ , \label{AH1} \\
\ell d(e^{-2A} y^{-1} j - y^{-2} \R w \wedge \I w) - F_0 H &=0 \ , \label{AH2} \\
d(e^{2A-\phi}\fpm a \omega) + 2 e^{-\phi} (- y \R w + i e^{2A} \I w ) \wedge a \omega &= 0 \label{Aomega} \ .
\end{align}
\end{subequations}
We can combine the first two of the above equations so as to obtain one which does not involve the NS--NS field strength $H$:
\begin{align}\label{Aj}
d\left[ (F_0^2 e^{2A} y + \ell^2 e^{-2A} y^{-1}) j - (F_0^2 y^2 + \ell^2 y^{-2}) \R w \wedge \I w \right] = 0 \ .
\end{align}
As pointed out earlier, $F_0$ and $\ell$ cannot be simultaneously zero, and hence when either of the two is, it follows
from \eqref{AH1} and \eqref{AH2} that $H=0$.

We will proceed by making a $2+4$ split of the internal manifold, with coordinates $\{y,\,\psi\}$ on the two-dimensional subspace. The differential operator is decomposed as
\begin{equation}
d = dy \wedge \partial_y + d\psi \wedge \partial_\psi + d_4 \ ,
\end{equation}
and the metric takes the form
\begin{equation}\label{ClassAM6}
ds^2_{M_6} = \frac{1}{e^{2A} - e^{-2A}y^2} dy^2 + \frac{1}{4}(e^{2A} - e^{-2A}y^2)(d\psi+\rho)^2 + g^{(4)}_{ij}(y,x^i) dx^i dx^j \ ,
\end{equation}
with $x^i$, $i=1,2,3,4$ coordinates on the four-dimensional subspace, $M_4$.

We can now decompose the three-form equations \eqref{Aj} and \eqref{Aomega}:
\begin{alignat}{9}\label{Asystem1}
\partial_\psi j &{}= 0 \ , & \qquad &
\ \ \ \partial_y(y^{-1} |\hat{a}|^2 j) &{}={}&  \frac{1}{2} (F_0^2 y^2 + \ell^2 y^{-2}) d_4 \rho \ , & \qquad &
\ \ \ d_4 (y^{-1} |\hat{a}|^2 j) &={}& 0 \ , \\
\partial_\psi \omega &{}=  - i \omega \ , & \qquad &
\partial_y (\fpm \hat{a} \omega) &{}={}& - 2 \frac{e^{-2A}y}{\fpm} \hat{a} \omega \ , & \qquad &
d_4 (\fpm \hat{a} \omega) &{}={}& - i \rho \wedge \fpm \hat{a} \omega \ , \nn
\end{alignat}
where we have introduced
\begin{equation}
\hat{a} \equiv e^{2A-\phi} a = -F_0 e^A y + i e^{-A} \ell \ .
\end{equation}

To further analyze the above equations it is convenient to rescale the data of the four-dimensional base as follows:
\begin{equation}
j = y |\hat{a}|^{-2} \hat \jmath \ , \qquad \omega = y |\hat{a}|^{-2} e^{-i(\psi+\theta)} \hat \omega \ , \qquad g^{(4)}_{ij} = y |\hat{a}|^{-2} \hat{g}^{(4)}_{ij} \ ,
\end{equation}
where $\theta \equiv \arg(\hat{a})$. Then \eqref{Asystem1} becomes
\eq{\label{Asystem2}
\begin{alignedat}{8}
\partial_\psi \hat{\jmath} &{}= 0 \ , & \qquad &
\partial_y \hat{\jmath} &{}={}& \frac{1}{2} (F_0^2 y^2 + \ell^2 y^{-2}) d_4 \rho \ , & \qquad &
d_4\hat{\jmath} &{}={}& 0 \ , \\
\partial_\psi \hat\omega &{}= 0 \ , & \qquad &
\partial_y \hat{\omega} &{}={}& - \frac{1}{2} (F_0^2 y^2 + \ell^2 y^{-2}) T \hat\omega \ , & \qquad &
d_4 \hat{\omega} &{}={}& i \hat{P} \wedge \hat\omega \ ,
\end{alignedat}
}
where
\sub{
\hat{P} &\equiv - \rho + i \frac{2 e^{4 A} (\ell^2 + F_0^2 y^4) }{(e^{4A}-y^2)(\ell^2 + F_0^2  e^{4A}  y^2)} d_4 A \ , \label{RicciCon} \\
T &\equiv \frac{\partial_y(e^{4A}y^2)}{(e^{4A}-y^2)(\ell^2 + F_0^2  e^{4A}  y^2)} \label{T} \ .
}
The last condition, $d_4 \hat{\omega} = i \hat{P} \wedge \hat\omega$, suggests that the almost complex structure defined by $\hat{\omega}$ is independent of $\psi$ and $y$ and integrable on the four-dimensional subspace $M_4$,  i.e.\ the latter is a complex manifold.
In addition, $d_4 \hat{\jmath} = 0$, and thus $\hat{g}^{(4)}$ is a family of K\"{a}hler metrics parametrized by $y$. Furthermore, $\hat{P}$ is the canonical Ricci form connection defined by the K\"{a}hler metric with the Ricci form $\hat{\mathfrak{R}} = d_4 \hat{P}$.

It is worth noting the similarity of the SU(2) structure we are studying here with the one that characterizes $\mathcal{N}=1$ supersymmetric
AdS$_5$ solutions of M-theory, studied in \cite{gauntlett-martelli-sparks-waldram-M}.\footnote{A similar resemblance occurs in the study of AdS$_3$ solutions\cite{Couzens:2017nnr}.} This close resemblance allows us to draw upon
certain results of the latter reference.

There are certain identities and conditions that derive from the system \eqref{Asystem2}, to which we now turn. The equation for $\partial_y \hat{\omega}$ determines the dependence of the volume of $M_4$ on $y$:
\begin{equation}
\partial_y \log \sqrt{\hat{g}^{(4)}} = (F_0^2 y^2 + \ell^2 y^{-2}) T \ .
\end{equation}
Given that the complex structure is independent of $y$ the following identity holds:
\begin{equation}
(\partial_y \hat{\jmath})^+ = \frac{1}{2} \partial_y \log \sqrt{\hat{g}^{(4)}} \, \hat{\jmath} \ ,
\end{equation}
where a plus superscript denotes the self-dual part of a two-form on $M_4$. Combining with the second equation of \eqref{Asystem2} we arrive at
\begin{equation}\label{d4rhoplus}
(d_4 \rho)^+ = - T \hat{\jmath} \ .
\end{equation}
Finally, the restrictions below hold as consequences of \eqref{Asystem2}:
\begin{equation}\label{Aint}
d_4 \rho \wedge \hat{\omega} = 0 \ , \qquad \left[\frac{8y}{(e^{2A}-e^{-2A}y^2)^2} d_4 A + i \partial_y \rho \right] \wedge \hat{\omega} = 0 \ .
\end{equation}

The four- five- and six-form parts of \eqref{S1}--\eqref{S5} are automatically satisfied given the conditions we have derived so far.

Let us now look at the rest of the fields. The dilaton is determined by \eqref{Aa} and the condition $|a|^2=1$ descending from \eqref{abc}:
\begin{equation}
e^{2\phi} = \frac{e^{2A}}{y^2 F_0^2 + e^{-4A} \ell^2} \ .
\end{equation}
The NS--NS field strength $H$ is given by either \eqref{AH1} or \eqref{AH2}, and its Bianchi identity $dH = 0$ is manifestly satisfied.
The R--R fields are determined by \eqref{F2} and are given by the expressions
\begin{subequations}
\begin{align}
F_2 &= \ell (\alpha_2 + e^{-2A} y^{-1} j) \ , \\
F_4 &= -F_0 (e^{2A} y j + \tfrac{1}{2} y^2 dy \wedge D\psi) \wedge \alpha_2 - \frac{1}{2} F_0 j \wedge j \ , \\
F_6 &= - 3 \ell e^{-4A} \vol_6 \ ,
\end{align}
\end{subequations}
where we have introduced the auxiliary two-form
\begin{equation}
\alpha_2 \equiv - \frac{1}{2} d \left(e^{-4A} y (d\psi+\rho)\right) + \frac{1}{2}  y^{-1} d\rho
\end{equation}
and $\vol_6 = \R w \wedge \I w \wedge \frac{1}{2} j \wedge j$.

For future reference, let us also note the following $B$-twisted fluxes $\tilde{F} \equiv e^{-B} F$, which will play a role when we examine flux quantization. This necessitates differentiating between the cases with the constants $\ell$, $F_0$ either generic or vanishing. We will consider the twisted fluxes only for the generic case with $F_0 \neq 0$, $\ell \neq 0$.
Local expressions for the NS--NS potential $B$ are easily read off from \eqref{AH1} or \eqref{AH2},
\sub{
B_1 &= - \frac{F_0}{\ell} \left(e^{2A} yj + \frac12 y^2 dy \wedge D\psi \right) \ ,\\
B_2 &= \frac{\ell}{F_0} \left( e^{-2A} y^{-1} j + \frac12 y^{-2} dy \wedge D\psi \right) \ .
}
While $B_2$ leads to shorter expressions for the remaining potentials, it has a singularity at $y=0$; we will thus work with $B_1$. We thus find the following expressions
\begin{subequations}\label{twistflux}
\begin{align}
\tilde{F}_2
&=
\frac{\ell}{2} d  \left(\left(y^{-1}-e^{-4A} y \right) D\psi\right) - F_0 (B_1- B_2) \ , \\
\tilde F_4
&= \frac{1}{2} \frac{F_0}{\ell^2}
\left(\frac{e^{4A}y^2}{F_0^2 e^{4A} y^2 + \ell^2} \hat{\jmath} \wedge \hat{\jmath} +  y^2 dy \wedge D\psi \wedge \hat{\jmath}\right) \ , \\
\tilde F_6
&= \frac{y^2}{4\ell^3} \frac{F_0^2 e^{4A} y^2 + 3 \ell^2}{F_0^2 e^{4A} y^2 + \ell^2} dy \wedge D\psi \wedge \hat{\jmath} \wedge \hat{\jmath} \  ,
\end{align}
\end{subequations}
where explicitly $\tilde F_4 \equiv F_4 - B_1 \wedge F_2 + \frac{1}{2} F_0 B_1 \wedge B_1$ and $\tilde{F}_6 \equiv F_6 - B_1 \wedge F_4 + \frac{1}{2} B_1^2 \wedge F_2 - \frac{1}{6} F_0 B_1^3$.

We will come back to these expressions in section \ref{cohom1}.

\subsection{Class HK}
In this section we look at the case $b\neq 0$. We find that in contrast to Class K this class of solutions is rather restricted and determined
up to constant parameters.

The condition \eqref{X1}, $\I(\xi) = 0$, fixes $z_3 = -1$, while \eqref{X2} gives $d(e^Af) = - \sqrt{f_+f_-} \R w$.
Similar to section \ref{sub:K}, we define the coordinate $y \equiv e^A f$ by
\begin{equation}
\R w = -\frac{1}{\sqrt{f_+f_-}} dy \ ,
\end{equation}
where once again $f_+f_- = e^{2A} - f^2 = e^{2A} - e^{-2A} y^2$, and a coordinate $\psi$ such that
$\xi = 4 \partial_\psi$.
From \eqref{xidual} it follows that
\begin{equation}
\I w = \frac{1}{2} \sqrt{f_+f_-} (d\psi + \rho) \ ,
\end{equation}
where $\rho$ is a one-form on the four-dimensional subspace orthogonal to $w$.

The zero-form component of \eqref{F2} yields $\R a = - e^{-A+\phi} y F_0$ while the one-form component of \eqref{S2} gives
$d(e^{3A-\phi} \I a) = 0$.
We can thus write
\begin{equation}
a = e^{-A+\phi}(- y F_0 + i e^{-2A} \ell), \qquad \ell = {\rm constant} \ .
\end{equation}
So far things are akin to section \ref{sub:K}.

From now on, however, analysis of the rest of the equations \eqref{SUSYeq} puts strong constraints on the SU$(2)$ structure and the functions that determine
the solution. We find:\footnote{In particular, these constraints are implied by the remaining one-form constraint and the three-form constraints. The two-, four-, five- and six-form constraints are trivially satisfied.}
\begin{equation}
\rho = d\varphi \ , \qquad dj = 0 \ , \qquad d(e^{i(\psi+\varphi)}\omega) = 0 \ ,
\end{equation}
where $\varphi$ is defined via $z_2 = e^{i(\varphi+\psi)}$ and satisfies $\partial_y \varphi = 0 = \partial_\psi \varphi$. We thus conclude
that the four-dimensional base of the internal manifold is hyper-K\"{a}hler\footnote{The phase $e^{i(\psi+\varphi)}$ can be absorbed in $\omega$.} and its metric is independent of $y$. Also, the connection of the fibration of the U$(1)$ isometry generated by $\xi$ over the base is flat. Furthermore, for the warp factor and the dilaton we find
\begin{equation}
e^{A} = L y^{-1/2} \ , \qquad e^{\phi} = g_s y^{-3/2} \ ,
\end{equation}
where $L$ and $g_s$ are constants. $b$ is also constant and is fixed by the relation $|a|^2 + b^2 = 1$ which becomes
$L^{-2} g_s^2 (F_0^2 + L^{-4} \ell^2) + b^2 = 1$.

The metric on the internal manifold, after a coordinate transformation $y = L \cos^{1/2}(\alpha)$, reads
\begin{equation}
ds^2_{M_6} = \frac{1}{4} e^{2A} (d\alpha^2 + \sin^2(\alpha)D\psi^2) + ds^2_{\rm HK}(x), \qquad e^{2A} = \frac{L}{\cos^{1/2}(\alpha)} \ .
\end{equation}
Here $D\psi \equiv d\psi + \rho$ and $ds^2_{\rm HK}(x)$ is the line element on the hyper-K\"{a}hler base, with $x$ denoting its coordinates.

Turning to the rest of the fields, the NS--NS field strength $H$ is zero, while the R--R fields can be read from \eqref{F2}. Their expressions
are:\footnote{The $\omega$ appearing here is the one following the redefinition $e^{i(\psi+\varphi)}\omega \to \omega$.}
\begin{subequations}
\begin{align}
F_2 &= -\frac12 \frac{\ell}{L} d\left(\cos^{3/2}(\a)\right) \wedge D \psi + \frac{\ell}{L^2} j + L e^{- \phi_0} b  \R \omega \ , \\
F_4 &= \frac12 L d\left(\cos^{3/2}(\a)\right) \wedge D \psi \wedge \left( F_0 j - L e^{- \phi_0} b  \I \omega \right) - \frac12 F_0 j \wedge j \ , \\
F_6 &= -3 \frac{\ell}{L^2} \cos \a  \vol_6 \ ,
\end{align}
\end{subequations}
where $\vol_6 = \R w \wedge \I w \wedge \frac{1}{2} j \wedge j$.

\section{Class K: complex Ansatz}\label{cohom1}
In this section we explore an Ansatz for the Class K of solutions consisting of
\begin{equation}\label{Ansatz}
d_4 A = 0 \ , \qquad \partial_y \rho = 0 \ .
\end{equation}
This Ansatz is equivalent to requiring that the holomorphic three-form
$\Omega = w \wedge \omega$ that characterizes the SU$(3)$ structure on the internal space $M_6$ satisfies $d\Omega = V \wedge \Omega$
for a one-form $V$, which in turn is equivalent to requiring that $M_6$ is a complex manifold.

From \eqref{Ansatz} and the definition \eqref{RicciCon} it follows that $\rho = - \hat{P}$. The Ricci form $\hat{\mathfrak{R}} = d_4 \hat{P}$ then reads $\hat{\mathfrak{R}} = - d_4 \rho$, and the second equation of \eqref{Asystem2} and \eqref{d4rhoplus} can be rewritten as
\begin{subequations}
\begin{align}
\hat{\mathfrak{R}} &= - \frac{2}{F_0^2 y^2 + \ell^2 y^{-2}} \partial_y \hat{\jmath} \label{ricci} \ , \\
\hat{\mathfrak{R}}^+ &= T \hat{\jmath} \label{self-ricci} \ .
\end{align}
\end{subequations}
From the above it can be inferred \cite{gauntlett-martelli-sparks-waldram-M} that the Ricci tensor on $M_4$, at fixed $y$, has two pairs of constant eigenvalues. For compact $M_4$, which is the case of interest, we can invoke \cite{apostolov-draghici-moroianu} stating (under the assumption that the Goldberg conjecture is true) that a compact K\"{a}hler four-manifold whose Ricci tensor has two distinct pairs of constant eigenvalues is locally the product of two Riemann surfaces of constant curvature. If the two pairs of eigenvalues are the same, then by definition the manifold is K\"{a}hler--Einstein. There are thus two classes to consider: either $M_4$ is K\"{a}hler--Einstein or is the product of two Riemann surfaces.

\subsection{K\"{a}hler--Einstein base}\label{sec:ke}
In this class
\begin{equation}\label{KEbase}
\hat{\mathfrak{R}} = \frac{\kappa}{Q(y)} \hat{\jmath} \ .
\end{equation}
with $\kappa = 0$ or $\kappa =  \pm 1$. The case $\kappa = 0$, corresponds to $M_4$ being hyper-K\"{a}hler and turns out to be the $b=0$
limit of the Class HK of solutions we examined in the previous section. We will thus restrict to $\kappa = \pm 1$.
The dependence of the metric of $M_4$ on $y$ is given by
\begin{equation}
\hat{g}^{(4)}(y,x^i) = Q(y) g_{\KE}(x^i)  \ ,
\end{equation}
where $g_{\KE}$ is a K\"{ahler}--Einstein metric of constant curvature $R = 4 \kappa$.

When combined with \eqref{ricci}, and the fact that $\partial_y \hat{\mathfrak{R}} = - d_4 \partial_y \rho = 0$ which is
part of the Ansatz \eqref{Ansatz}, the condition \eqref{KEbase} fixes
\begin{equation}\label{KEQ}
\frac{Q}{\kappa} = \frac{1}{6} (3 \ell^2 y^{-1} - F_0^2 y^3) + \nu \ , \qquad \nu = \text{constant} \ .
\end{equation}
In combination with \eqref{self-ricci}, \eqref{KEbase} gives the ordinary differential equation (ODE) $T = \kappa/Q$, which determines the warp factor $A$. Given the expression \eqref{T} for $T$ and defining
\begin{equation}
p(y) \equiv e^{4A} y^2~,
\end{equation}
this becomes a Riccati:
\begin{equation}\label{eq:riccati}
y^2 \frac{Q}{\kappa} \frac{dp}{dy} =  F_0^2 p^2 + (\ell^2-F_0^2 y^4) p - \ell^2 y^4 \ .
\end{equation}
We were able to solve this Riccati equation analytically:
\begin{equation}\label{KEsol}
p = \ell^2 y^2 \frac{3 \ell^2 \mu - 9 \ell^2 y^2  - 12 \nu  y^3 +   F_0^2 y^6}{ 9 \ell^4 + 36 \ell^2 \nu y + (36 \nu^2 - 3 \ell^2 F^2_0 \mu) y^2 + 3 \ell^2 F_0^2 y^4} \ ,
\end{equation}
where $\mu$ is a constant parameter. Note that in this parametrization the limit $\ell \to 0$ is not well-defined since the solution
becomes trivial. The $\ell \to 0$ limit is well-defined after shifting $\mu \to 12 \nu^2/(F^2_0 \ell^2) + \mu$.

\subsubsection{Regularity and boundary conditions} 
\label{sub:reg}

We now turn to the analysis of the geometry of the solutions, which we will carry out in terms of a rescaled coordinate $ x \propto y$. We will specify the constant rescaling factor later on, for the cases  (i)  $F_0 \neq 0$ and $\ell \neq 0$ (generic),  (ii) $\ell = 0$, and (iii) $F_0 = 0$, separately.

The metric \eqref{ClassAM6} on the internal manifold takes the form:
\begin{subequations}\label{eq:ke}
\begin{equation}\label{eq:ke-met}
e^{-2A} ds^2_{M_6} = - \frac14 \frac{q'}{x q} dx^2 - \frac{q}{x q' - 4 q} D\psi^2 + \frac{\kappa q'}{3 q' - x q''} ds^2_{\KE} \  ,
\end{equation}
where $q=q(x)$ is a polynomial (of degree 6 if $F_0\neq0$), and a prime denotes differentiation.  The warp factor is given by
\begin{equation}\label{eq:ke-A}
L^{-2} e^{2A} = \sqrt{\frac{x^2q'-4xq}{q'}} \ .
\end{equation}
The dilaton is given by
\begin{equation}\label{eq:ke-phi}
	g_s^{-2} e^{2 \phi} = \frac{x q'}{(3q'-x q'')^2} \left(\frac{x^2q'-4xq}{q'} \right)^{3/2} \ .
\end{equation}
\end{subequations}
$L$ and $g_s$ are two integration constants which we will specify in terms of the constants appearing in $p$ later on.

Positivity of the metric and the dilaton requires
\begin{equation}\label{eq:pos}
	q<0 \ , \qquad x q'>0 \ ,\qquad \kappa (3 x q' - x^2 q'')>0 \ .
\end{equation}
These conditions will only be realized on an interval of $x$. What happens to $q$ at an endpoint $x_0$ of this interval dictates the physical interpretation of the solution around that point. We summarize our conclusions in Table \ref{tab:bc}. For example, we see from there that if $q$ has a simple zero at a point $x_0\neq0$, the $S^1$ parametrized by $\psi$ shrinks in such a way as to make the geometry regular, provided that the periodicity $\Delta \psi$ is chosen to be $2\pi$. If this happens at both endpoints of the interval, the solution is fully regular.

\begin{table}
\begin{center}
\begin{tabular}{ccccc}
$x_0$ &$q(x_0)$ & $q'(x_0)$ & $q''(x_0)$ & interpretation \\\hline\hline
 &$0$ &  & & regular \\\hline
 & & $0$ & & O4 \\\hline
 &$0$ & $0$ & $0$ & conical CY\\\hline
$0$ & & $0$ & $0$ &O8 \\\hline
\end{tabular}
\end{center}
\caption{Various boundary conditions for the polynomial $q$ at an endpoint $x_0$, and their interpretation. Empty entries are meant to be non-zero.}
\label{tab:bc}
\end{table}

Here are some details about each of these cases.
\begin{itemize}

\item[] \textbf{Simple zero: regular endpoint.} Near a simple zero $x=x_0$ of $q$, the warp factor and dilaton go to constants, while the internal metric behaves as
\begin{equation}\label{eq:loc-zero}
	\frac{1}{L^2|x_0|} ds^2_{M_6} \sim  \frac1{x_0} \left(\frac14\frac{dx^2}{x_0-x} + (x_0-x)D \psi^2\right) + \frac{\kappa q'_0}{3 q'_0 - x_0 q''_0}ds^2_{\KE} \ ,
\end{equation}
where $q'_0 \equiv q'(x_0)$, $q''_0 \equiv q''(x_0)$ ($x_0$ will never be zero). Positivity of the metric requires that if $x_0>0 \Rightarrow x<x_0$, and if $x_0<0 \Rightarrow x>x_0$. Choosing for definiteness the first case, and introducing $r=\sqrt{x_0-x}$ we see that the parenthesis in (\ref{eq:loc-zero}) becomes $dr^2 + r^2 D \psi^2$, which is the metric of $\mathbb{R}^2$ (fibred over the K\"{a}hler--Einstein base), with the condition that the periodicity of $\psi$ is  taken to be $\Delta \psi= 2\pi$.

\item[] \textbf{Extremum: O4-plane.} At a point $x_0\neq 0$ where $q'(x_0)=0$, the ten-dimensional metric and dilaton behave as
\begin{align}\label{eq:ext-o4}
	\frac1{2L^2}\sqrt{-\frac{q''_0}{x_0 q_0}} ds_{}^2 &\sim
	\frac1{\sqrt{x-x_0}}\left(ds^2_{\mathrm{AdS}_4} + \frac14 D \psi^2 \right) + \sqrt{x-x_0}\left(-\frac{q''_0 dx^2}{4x_0q_0}   - \frac\kappa{x_0} ds^2_{\KE} \right) \ , \\
	g_s^{-2} e^{2\phi} &\sim \frac{1}{x_0 q''_0}\left(\frac{-4x_0q_0}{q''_0}\right)^{3/2}\frac{1}{\sqrt{x-x_0}} \ ,	
\end{align}
where $q_0\equiv q(x_0)$. Positivity of the metric and the dilaton requires that if $x_0 q''_0>0 \Rightarrow x>x_0$, and if $q''_0 x_0<0 \Rightarrow x<x_0$ (in the above equation we have recorded the first case). It also requires $\kappa q''_0<0$. One recognizes the usual structure $H^{-1/2} ds^2_{\parallel}+ H^{1/2}ds^2_\perp $ for extended objects, with $H\sim x-x_0$. Since there are five parallel directions, this signals the presence of a four-dimensional object; the fact that the function is linear matches with the behavior of an O4-plane near the point where its harmonic function goes to zero. The dilaton matches the behavior $e^{2\phi} \propto H^{(3-p)/2}$ of an O$p$-plane again for $p=4$. Thus we conclude that this singularity corresponds to the presence of an O4-plane extended along AdS$_4$.

We should also point out, however, that the local structure of the singularity does not clarify if the orientifold is smeared over $\KE$. Suppose one places a fully localized O4-plane at the tip of a cone $C(Y_4)$ of metric $dx^2+x^2 ds^2_{Y_4}$. Near the tip, the backreacted metric is then of the form
\begin{equation}\label{eq:O4}
	ds^2 = H_{\mathrm{O4}}^{-1/2} ds^2_{\parallel} + H_{\mathrm{O4}}^{1/2} (dx^2 + x^2 ds^2_{Y_4}) \ ,\qquad H_\mathrm{O4} = 1- \left(\frac{x_0}{x}\right)^3 \ .
\end{equation}
On the other hand, an O4-plane that is partially smeared along a four-dimensional manifold $Y_4$ would have a metric
\begin{equation}\label{eq:smO4}
	ds^2 = H_\mathrm{smO4}^{-1/2} ds^2_{\parallel} + H_\mathrm{smO4}^{1/2} (dx_9^2 + ds^2_{Y_4}) \ ,\qquad H_\mathrm{smO4} = a+ b|x_9| \ ;
\end{equation}
since $H_\mathrm{smO4}$ is now a harmonic function of one dimension only, it is piecewise linear.

The metric (\ref{eq:O4}) ceases to make sense at $x=x_0$. Expanding around this point, $H_\mathrm{O4}\sim (x-x_0)$, and we would obtain (\ref{eq:ext-o4}) (up to constants that can be reabsorbed). On the other hand, (\ref{eq:smO4}) for $a=0$ also gives (\ref{eq:ext-o4}), upon identifying $x-x_0=|x_9|$. In this sense, it is not entirely clear if (\ref{eq:ext-o4}) should be considered as smeared over $\KE$ or not. (Such considerations also apply to O$p$-planes for $p\neq 4$.) For $Y_4=S^4$, (\ref{eq:O4}) is the simplest interpretation; for $Y_4$ a K\"{a}hler-Einstein manifold, the singularity $C(Y_4)$ would be bad (in that for example it would not be Ricci-flat, as one would expect before placing an object on it), and (\ref{eq:smO4}) seems the simplest interpretation. We thus conclude (\ref{eq:ext-o4}) is an O4-plane that is smeared over $\KE$.

Of course smeared orientifolds have rather limited physical validity; nevertheless, for completeness, we will include them in our analysis.

\item[] \textbf{Triple zero: conical Calabi--Yau singularity.}  Near a triple zero $x=x_0$ of $q$, the warp factor and dilaton go to constants, while the internal metric behaves as
\begin{equation}\label{eq:concy}
	\frac{1}{L^2|x_0|} ds^2_{M_6} \sim \frac{3}{x_0} \left[ \frac14\frac{dx^2}{x_0-x} + (x_0-x)\left(\frac1{9} D \psi^2 + \frac\kappa{6} ds^2_{\KE} \right) \right ] \ .
\end{equation}
Positivity works as in the case of a simple zero, but it also requires $\kappa = 1$. Upon introducing $r=\sqrt{x_0-x}$, (\ref{eq:concy}) becomes proportional to
\begin{equation}\label{eq:concyr}
	dr^2 + r^2\left(\frac19 D \psi^2 + \frac  1 6 ds^2_{\KE} \right) \ .
\end{equation}
The metric in parenthesis is a regular Sasaki--Einstein metric, built as U(1) bundle over the K\"{a}hler--Einstein base ($\KE$). Thus (\ref{eq:concyr}) represents a conical Calabi--Yau singularity. In the particular case that $\KE$  is $\mathbb{CP}^2$, this is in fact an orbifold singularity.

We can be a little more precise. $d(D \psi)$ is the Ricci form of $\KE$, which in de Rham cohomology represents the first Chern class $c_1$. The integral of the latter over the two-cycles $\CC_i$ of $\KE$ is $2\pi n_i$, $n_i \in \mathbb{Z}$. If the periodicity  of the $S^1$ coordinate is $\Delta \psi=2\pi$, the total space is the U(1) bundle associated to the canonical bundle over $\KE$. In that case the conical singularity (\ref{eq:concyr}) is the complex cone over $\KE$. If $\mathrm{gcd}(n_i)\ge 1$, there is also the possibility of taking the periodicity to be $2\pi \times \mathrm{gcd}(n_i)$.

For example, if $\KE=\mathbb{CP}^2$,  there is only one cycle $\CC$ with $n=3$; taking $\Delta \psi =2 \pi$ gives the orbifold singularity
\begin{equation}\label{eq:z3}
	\mathbb{C}^3/\mathbb{Z}_3 \ ,
\end{equation}
but one can also consider $\Delta \psi = 6 \pi$, for which (\ref{eq:concyr}) in fact becomes the fully regular space $\mathbb{C}^3$. This possibility is not available if at the other endpoint one has a single zero, where $\Delta \psi$ is necessarily $2\pi$. However, as we will see, in one case there is a triple zero is present at both endpoints, and in that case $\Delta \psi=6 \pi$ is possible. This will correspond to the Guarino--Jafferis--Varela (GJV) solution \cite{guarino-jafferis-varela}.

If $\KE=\mathbb{CP}^1\times \mathbb{CP}^1$, there are two $\CC_i$ with $n_i=2$. With $\Delta \psi=2\pi$, (\ref{eq:concyr}) becomes a $\mathbb{Z}_2$ quotient of the conifold singularity; one also has the possibility of taking $\Delta \psi=4\pi$, for which one obtains the original conifold singularity. Again this option is only available if a triple zero appears at both endpoints. We will see later, when considering the product base class in section \ref{sec:pro-base}, that this case has in fact a richer array of possibilities.

\item[] \textbf{Inflection point at the origin: O8-plane.} When $q'(0)=q''(0)=0$, around $x=0$, the metric and dilaton behave as
\begin{align}\label{eq:o8}
	\frac1{L^2}\sqrt{-\frac{q_3}{8q_0}}ds^2_{10} &\sim \frac1{\sqrt x} \left(ds^2_{\mathrm{AdS}_4}
	+ \frac14 D \psi^2 + \kappa ds^2_{\KE}\right) -\frac{q_3}{8 q_0} \sqrt{x} \, dx^2 \ , \\[5pt]
          g_s^{-2} e^{2\phi} &\sim \frac{2}{q_3} \left(-\frac{8q_0}{q_3} \right)^{3/2} x^{-5/2} \ .
\end{align}
where $q_0\equiv q(0)$ and $q_3 \equiv q^{(3)}(0)$. This time we recognize an O8-plane localized at $x=0$.
\end{itemize}

\subsubsection{Generic case} 
\label{sub:gen}
We now turn to examining the parameter space of solutions for the generic case, by which we mean $F_0\neq 0$ and $\ell \neq 0$.
The rescaling of the coordinate $y$ that we mentioned at the beginning of \ref{sub:reg} is
\begin{equation}
x= \sqrt{\frac {F_0}{\sqrt 3 \ell}} y \ .
\end{equation}
We will also rescale the constant parameters that appeared in \eqref{KEsol}  and introduce
\begin{equation}\label{eq:beta-gamma}
\beta = \frac{F_0}{2\sqrt{3} \ell} \mu \ , \qquad
\gamma = \frac{1}{\ell^2} \sqrt{\frac{\sqrt{3} \ell}{F_0}} \nu \ .
\end{equation}
The polynomial $q$ that determines the solution then reads
\begin{equation}\label{KEgenericq}
q = x^6 + 3 (2\gamma^2 - \beta) x^4 + 8 \gamma x^3 + 3 x^2 - \beta \ ,
\end{equation}
while the constants $L$, $g_s$ introduced in the previous section are:
\begin{equation}\label{eq:Lgsgen}
	L^4 = \frac{\sqrt 3 \ell}{F_0} \ , \qquad g_s^2= \frac{3^{7/4}\cdot 8}{\sqrt{\ell F_0^3}} \ .
\end{equation}

For later use, we note that the Riccati ODE (\ref{eq:riccati}) implies an ODE directly on the polynomial $q$:
\begin{equation}\label{eq:qode}
	\frac1{24}(3q'-xq'')^2 = (xq'-4q)(1+3x^4)+4q\,.
\end{equation}
It is also useful to notice that in this case
\begin{equation}
	3q'-xq''= 12 x (1 +2 \gamma x - x^4) \ .
\end{equation}

We now have to study for which values of the parameters $\beta$, $\gamma$ the positivity conditions (\ref{eq:pos}) are satisfied. First of all, in order for $q<0$ we need to require $\beta>0$. We then have to subdivide this region according to the nature of the zeros of $q$, and the presence of maxima or minima. To identify these subregions, it is useful to look at the discriminant of $q$:
\begin{equation}
	\Delta(q) = 2^{10} 3^6 \left(\beta^4 - 6 \gamma^2 \beta^3 - (2-12\gamma^4) \beta^2 - \gamma^2(10+8\gamma^4)\beta + \gamma^4 + 1 \right)^2 \beta \ .
\end{equation}
For every $\beta$, $\Delta(q)=0$ has two solutions $\beta=\beta_\pm$, $\beta_-\le \beta_+$. (This can be seen by considering the discriminant of $\Delta(q)$ with respect to $\beta$, which is always negative.) Notice that $\beta \Delta(q')^2 \propto \Delta(q)$, and that res$(q,q'')$ divides $\Delta(q')$; this implies a double zero is in fact also a triple zero. This also follows from (\ref{eq:qode}).

\begin{figure}[ht]
	\centering
		\includegraphics[height=3in]{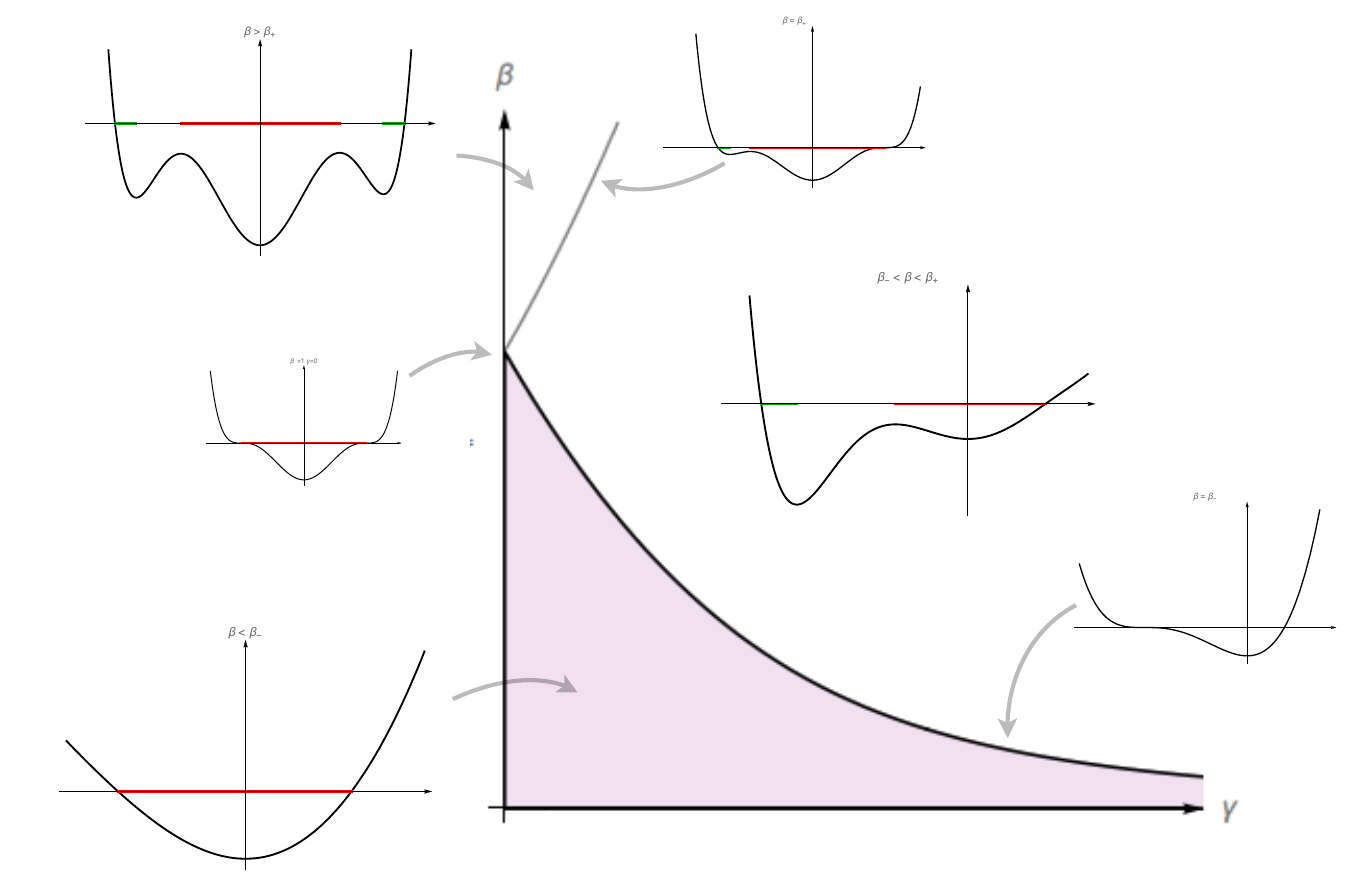}
	\caption{Plots of $q(x)$ corresponding to various regions of parameter space.}
	\label{fig:KE-plots}
\end{figure}

There are six different cases, which we discuss in turn and describe in figure \ref{fig:KE-plots}. For definiteness, we will consider $\gamma \ge 0$; the discussion for $\gamma \le 0$ is similar.

\begin{itemize}
	\item For $\beta<\beta_-$, $q$ has two simples zeros $x_\pm$; in the interval $[x_-,x_+]$, the conditions (\ref{eq:pos}) are met with $\kappa=+1$. According to our discussion in section \ref{sub:reg}, at both simple zeros the $S^1$ circle shrinks in such a way that the solution is regular. Thus the internal space is smooth and the solution is fully regular.
	
	The solutions previously found numerically fall in this region. The first to be found were the ones in \cite{petrini-zaffaroni}, which should correspond to $\gamma=0$, $\beta \le 1$, with $\KE=\mathbb{CP}^2$. In \cite{lust-tsimpis-singlet-2} it was later suggested that the $\mathbb{CP}^2$ could be replaced by an arbitrary regular $\KE$. This was worked out explicitly in \cite{tz} for $\KE=\mathbb{CP}^1\times \mathbb{CP}^1$; our regular solutions corresponds to the $q_1=\tilde q_1$ slice in \cite[Fig.~3]{tz}. (We will see later how the rest of that figure is reproduced.)
		
	\item At $\beta=\beta_-$, the discriminant $\Delta(q)=0$, and as we remarked also $\Delta(q')=0$; so the simple zero $x_-$ becomes a triple zero. As we discussed in section \ref{sub:reg}, this means that one of the regular points becomes a conical Calabi--Yau singularity; if $\KE=\mathbb{CP}^2$, this is a $\mathbb{Z}_3$ singularity.
	\item For $\beta_- < \beta < \beta_+$, the triple zero $x_-$ splits into a single zero $x_-$, a local minimum $x_1$ and a local maximum $x_2$ ($x_-<x_1<x_2$). Now the positivity conditions (\ref{eq:pos}) are met for $\kappa=+1$ between the maximum and the zero: $x\in [x_2,x_+]$. However, a new possibility also appears: for $\kappa=-1$, the positivity conditions are met in the interval $x\in [x_-,x_1]$. Both of these correspond to solutions with a single O4-plane singularity.
	\item At $\beta=\beta_+$, the other simple zero $x_+$ becomes a triple zero. Thus the solution with $\kappa=+1$ develops a Calabi--Yau singularity, besides the O4-plane singularity it already had; the solution with $\kappa=-1$ still has a single O4-plane singularity.
	\item For $\beta > \beta_+$, the new triple zero splits into a local maximum $x_3$, a local minimum $x_4$ and a simple zero $x_+$. Now the $\kappa=+1$ interval is $x\in [x_2,x_3]$, and corresponds to a solution with two O4-plane singularities. Moreover there are two intervals that are allowed for $\kappa=-1$: $x\in [x_-,x_1]$ and $x\in[x_4,x_+]$. Both correspond to a solution with a single O4-plane singularity.
	\item Finally, at $\beta=1$, $\gamma=0$ we have that $\beta_-=\beta_+$. In this case the two triple zeros appear together; the allowed interval is between them and works for $\kappa=+1$. The solution has two Calabi--Yau singularities. When $\KE=\mathbb{CP}^2$, with $\Delta \psi=2\pi$ these are two orbifold singularities (\ref{eq:z3}). (This is the periodicity originally considered in \cite{petrini-zaffaroni}.) As we discussed there, in this case one can also take the periodicity of the $S^1$ to be $\Delta \psi=6\pi$, in which case the space becomes fully smooth again; this is the GJV solution \cite{guarino-jafferis-varela}.\footnote{The coordinate transformation that brings the solution to the form of \cite{guarino-jafferis-varela} is $x = \cos\alpha$ and $D\psi = 3 \eta$. The parameters $e^{\phi_0}$ and  $L$ of \cite{guarino-jafferis-varela} are identified as $e^{\phi_0} = 3^{-3/8} 2^{1/4} \ell^{-1/4} F_0^{-3/4}$ and $L^2 = 3^{-1/16} 2^{-5/8} \ell^{5/8} F_0^{-1/8}$. Note also that the solution in \cite{guarino-jafferis-varela} is in the Einstein frame, whereas we work in the string frame.}
For more general $\KE$, this solution was discussed in \cite{fluder-sparks}.
\end{itemize}

\subsubsection{Flux quantization for the generic case} 
\label{sub:fluxq}

Let us now discuss flux quantization for the generic solutions of section \ref{sub:gen}.

First we need to return to the $B$-twisted fluxes \eqref{twistflux} and introduce potentials $C_{k-1}$ such that $dC_{k-1} = \tilde F_k$.
Explicitly,
\begin{equation}\label{eq:C135}
C_1 = f_2 D \psi \ , \qquad
C_3 = f_4 D\psi \wedge j_{\rm KE} \ , \qquad
C_5 = f_6 D\psi \wedge j_{\rm KE} \wedge j_{\rm KE} \ ,
\end{equation}
where
\begin{subequations}\label{eq:f2f4f6}
\begin{align}
f_2 &= 2\ell\sqrt{\frac{F_0}{\sqrt 3 \ell}}\left(\frac1x\frac q{4q-xq'}-\frac{3q'-x q''}{48x^2}\right) \ , \\
f_4 &= \frac{\ell}{16\sqrt3} \kappa  (4q - x q') \ , \\
f_6 &= \frac{\ell}{16} \sqrt{\frac{\sqrt{3} \ell}{F_0}} \left[\frac{q'}{24}(3+x^4)-\frac q6 x^3+\frac x9\left(3+x^4+\frac{3q'-xq''}{8x}\right)^2\right] \ .
\end{align}
\end{subequations}
The fluxes $\tilde F_k$ are closed; they have been defined using a particular choice $B_1$ for the $B$ field. In fact it is also possible to add to it a closed two-form $b$, so that $B= B_1 + b$; this defines new fluxes
\begin{equation}\label{eq:Fb}
	\tilde F^b \equiv e^b \tilde F\,,
\end{equation}
which are also closed. Explicitly, $\tilde F_2^b \equiv \tilde F_2 + b F_0$, $\tilde F_4^b \equiv \tilde F_4 + b \tilde F_2 + \frac12 b^2 F_0$, $\tilde F_6^b \equiv \tilde F_6 + b \tilde F_4 + \frac12 b^2 \tilde F_2+ \frac16 b^3 F_0$. Flux quantization now imposes that the periods of these should be quantized, as well as that $2\pi F_0\equiv n_0 \in \mathbb{Z} $ (working in string units $l_s=1$). It constrains the parameters $\ell$, $\beta$, $\gamma$ of the solution, as well as the two-form $b$.

We will now work out more precisely what this implies for regular generic solutions. In particular, we will assume $\Delta \psi=2\pi$ and $\beta< \beta_-$, in the language of section \ref{sub:gen}; topologically, $M_6$ is an $S^2$-bundle over $\KE$.

The second homology of $M_6$ is given by the fiber $\CC_0\equiv S^2_{\psi,x}$ spanned by $\psi$ and $x$, and by the two-cycles $\CC_i$, $i=1,\ldots,h_2(\KE)$. More precisely, a lift of these two-cycles is given by a section of the fibration obtained by setting $x$ to one of the endpoints, say $x_+$. ($x_-$ would also work, but a random value would not define a cycle in $M_6$, because of the topological non-triviality of the fibration of the $\psi$ coordinate.) A basis for the cohomology $H^2$ can be taken to be $\omega^I$, $I=0,\ldots,h^2(\KE)$;  $\omega^0 \equiv d(s(x) D \psi)$, where $s(x)$  is a function which at the two simple zeros $x_\pm$ has second-order expansion $s\sim \pm (1 +(x-x_\pm)^2)+\ldots$, and $\omega^i$ are the elements of a basis for $H^2(\KE)$. We will expand the closed two-form $b$ in this basis: $b= b_I \omega^I$. Similarly, a basis of four-cycles can be obtained by the $\tilde \CC^i\equiv S^2_{\psi,x}\times \CC_i$ and by $\tilde \CC^0\equiv \KE$. Finally, the triple intersection form $d^{IJK}$ of $M_6$ will have non-zero entries $d^{0JK}= c^{JK}$, the intersection form of $\KE$.

We can now define the periods
\begin{equation}
	n_{2I}^b \equiv \frac1{2\pi}\int_{\CC_I} \tilde F_2^b \, ,\qquad n_{4}^{I\,b} \equiv \frac1{(2\pi)^3}\int_{\tilde \CC^I} \tilde F_4^b \, ,\qquad n_6^b \equiv \frac1{(2\pi)^3}\int_{M_6}\tilde F_6^b\,.
\end{equation}
The periods at $b=0$,  $n_{2I}\equiv n_{2I}^{b=0}$, $n_4^I\equiv n_{4}^{I\,b=0}$, $n_6\equiv n_6^{b=0}$, are computed more directly as integrals of the $\tilde F_k$. The two are related by the $b$-transform (\ref{eq:Fb}): this gives $n_{2I}^b = n_{2I}- b_I n_0$, $n_4^{I\,b} = n_4^I + d^{IJK} b_J\left(n_{2K}+\frac12 b_K n_0\right)$, $n_6^b= n_6+ b_I n_4^I +\frac12 d^{IJK}b_I b_J \left(n_{2K}+\frac13 b_K n_0\right)$. From \eqref{eq:C135}), \eqref{eq:f2f4f6} we can now compute the relevant integrals:
\begin{equation}\label{eq:ni}
\begin{split}
	&n_{20}= f_{2+} - f_{2-} \, ,\qquad n_{2i} = K_i f_{2+} \, ,\\
	n_4^i = \frac{K_i}{2\pi \kappa} (f_{4+}&- f_{4-}) \, ,\qquad n_4^0= -\frac{K^2}{2\pi \kappa}f_{4+} \, ,\qquad n_6 = \frac{K^2}{4\pi^2} (f_{6+}- f_{6-})\,,
\end{split}	
\end{equation}
where $f_{k\pm}\equiv f_k(x_\pm)$ the $K_i$ are the Chern class integers of the canonical bundle; $2\pi K_i$ are the integrals of the Ricci form over the two-cycles $\CC_i$ of the $\KE$. We also defined $K^2\equiv K_i K_j c^{ij}$.  (\ref{eq:ni}) can be further evaluated using the expressions for the $f_k$ in \eqref{eq:f2f4f6}. In doing so, it is useful to note that (\ref{eq:qode}) implies that at a single zero $x_0$ of $q$:
\begin{equation}
	(3 q_0'-x_0 q''_0)^2 = 24 q'_0 x_0 (1+3 x_0^4)\,.
\end{equation}
So in particular
\begin{equation}
\begin{split}
	&f_2(x_0)=-\ell\sqrt{\frac{F_0}{\sqrt3\ell}}\sqrt{\frac{q'_0(1+3x_0^4)}{24x_0^3}} \, ,\qquad f_4(x_0)=-\frac{\ell}{16\sqrt3} \kappa x_0 q'_0 \, ,\\
	&f_6(x_0)= \frac{\ell}{16}\sqrt{ \frac{\sqrt{3}\ell}{F_0} }(3+x_0^4)\left[\frac16q'_0-\frac{1}{3\sqrt6}\sqrt{x_0 q'_0 (1+3x_0^4)} +\frac{x_0}9(3+x_0^4)\right]\,.
\end{split}
\end{equation}

The $n_{2I}$ now determine $b_I = \frac1{n_0}(n_{2I}^b - n_{2I})$; one can then eliminate them from the remaining quantization conditions. A practical way of doing this is to introduce some combinations of the flux quanta that are invariant under $b$-transform $\tilde F\to \tilde F^b$, generalizing slightly results in \cite{ajtz}:
\begin{equation}\label{eq:I}
	I_4^I \equiv d^{IJK} n_{2J}n_{2K}- 2 n_0 n_4^I 	\, ,\qquad I_6 \equiv d^{IJK}n_{2I}n_{2J} n_{2K} +3 n_0^2 n_6 -3 n_0 n_{2I}n_4^I\,.
\end{equation}
(These come from the expansion in form basis of $\tilde F_2^2-2 F_0 \tilde F_4$ and $\tilde F_2^3+3 F_0^2 \tilde F_6 -3 F_0 \tilde F_2 \tilde F_4$.)
Indeed one can check that the $I_4^I$ and $I_6$ remain the same if one replaces $n_{2I}\to n_{2I}^b$, $n_4^I\to n_4^{I\,b}$, $n_6 \to n_6^b$. For us these invariants evaluate to
\begin{equation}
\begin{split}
	I^0_4&= K^2 \left(f_{2+}^2+\frac{n_0}{\pi k}f_{4+}\right) \, ,\qquad I^i_4=2K^i\left[-f_{2+}(f_{2+}-f_{2-})-\frac{n_0}{2\pi \kappa}(f_{4+}-f_{4-})\right] \\
	I_6 &= 3K^2 \left[f_{2+}^2(f_{2+}-f_{2-})+\frac{n_0}{2\pi \kappa}(2f_{4+}f_{2+}-f_{4+}f_{2-}+f_{4-}f_{2+})+\frac{3n_0}{4\pi^2}(f_{6+}-f_{6-})\right]\,.
\end{split}
\end{equation}
Once a set of flux quanta is specified, solving these equations will specify the parameters $\ell$, $\beta$, $\gamma$ of the solution.


\subsubsection{$\ell=0$} 
\label{sub:l0}

We will now examine the solutions with $\ell=0$. The rescaling of the coordinate $y$, appropriate for this case,  is
\begin{equation}
x= \frac y{y_0} \  , \qquad y_0\equiv \left(\frac{6 \nu}{F_0^2}\right)^{1/3} \  ,
\end{equation}
and we will also introduce
\begin{equation}
	\s = - \frac{3 \mu}{y_0^2} \ .
\end{equation}
The solution is then in the form (\ref{eq:ke}) with
\begin{equation}\label{eq:qLgsl0}
	q= x^6+ \frac{\s}{2} x^4 + 4 x^3 - \frac{1}{2} \, ,\qquad
\end{equation}
which gives
\begin{equation}
	3q'-xq''= -12 x^2(x^3-1) \ .
\end{equation}
The constants that appear in the warp factor and the dilaton are $L^2= |y_0|$ and $g^2_s = 72/(|y_0| F_0^2)$.

In this case, the analysis is easier, because there is only one parameter, $\s$, to vary. The discriminant of $q$ is $2(\s^3+9^3)^2$; thus there always two simple zeros, except for $\s=-9$, when one of the two becomes a triple zero. $q'$ always has a double zero at the origin, but the discriminant of $q'/x^2$ is $-192(\s^3+9^3)$ (so again we have $\Delta(q)\propto \Delta(q')^2$),  and the sign shows that there is a single extremum $x_1$ for $\s>-9$, and three extrema $x_1$, $x_2$, $x_3$ for $\s<-9$. In addition, there is a inflection point at $x=0$, which from section \ref{sub:reg} we know to correspond to an O8-plane.

We divide the analysis in three cases:

\begin{itemize}
	\item For $\s>-9$, between the two zeros $x_-<0$ and $x_+>0$, $q$ has a minimum at $x=x_1<0$ and the inflection point at $x=0$. The intervals where (\ref{eq:pos}) are realized are $x\in [x_-,x_1]$ with $\kappa=-1$, and $x\in [0,x_+]$ with $\kappa=+1$. The former corresponds to a solution with a single O4-plane singularity; the latter to a solution with a single O8-plane singularity.
	
	\item For $\s=-9$, the zero $x_+$ becomes triple; the allowed intervals remain the same as in the previous case, but the $\kappa=+1$ case now develops a Calabi--Yau conical singularity at $x=x_+$.
	
	\item For $\s<-9$, the zero $x_+$ splits in a maximum $x_2$, a minimum $x_3$ and a simple zero $x_+$ (all three greater than zero). There are now three allowed intervals: the old one $x\in [x_-,x_1]$, still for $\kappa=-1$, an interval $x\in[0,x_2]$ for $\kappa=+1$, and a new one $x\in[x_2,x_+]$ for $\kappa=-1$. These two new possibilites correspond to a solution with an O8-plane and an O4-plane singularity, and to a solution with a single O4-plane singularity, respectively.
\end{itemize}

\subsubsection{$F_0=0$} 
\label{sub:f00}
In this limit, the rescaling of the coordinate $y$ to the coordinate $x$ is
\begin{equation}
y = \frac{\ell^2}{\nu} x  \, .
\end{equation}
Furthermore the constant parameter $\mu$ is rescaled to $s$  as $\mu = 3 (\ell^4/\nu^2) s$.
From (\ref{eq:beta-gamma}) we see
\begin{equation}\label{eq:s-bg}
	s=\frac23 \beta \gamma^2\,.
\end{equation}
The massless limit is now obtained by taking $\beta\to 0$ with $s$ constant. We obtain (\ref{eq:ke}) with
\begin{equation}\label{eq:qLgsf00}
	q=x^4+\frac43 x^3+\frac12 x^2-\frac s4 \ ,\qquad L^2=\frac{\ell^2}{\nu} \ ,\qquad g^2_s=\frac{4\ell^4}{\nu^3} \ .
\end{equation}
These solutions uplift to M-theory AdS$_4 \times Y^{p,k}$ solutions, where $Y^{p,k}$ are the well-known Sasaki--Einstein seven-manifolds of \cite{gauntlett-martelli-sparks-waldram-SE7}. To see this, one has to perform the further change of coordinate
\begin{equation}
	x=\rho^2-\frac12 \ ,
\end{equation}
and set the constants in \cite[Sec.~2]{gauntlett-martelli-sparks-waldram-SE7} as $\{\Lambda, \kappa, \lambda\} = \{8,12s-1,2\ell/\nu\}$.

In the limit $s\to 0$, the interval of definition for $x$ shrinks to zero. However, one can define $S \equiv s/\nu^2$ and take $S\to 0$; in this limit the solution remains well-defined. After introducing the coordinate $\theta$ by $\cos \theta \equiv \sqrt{2/s} \, x$, it becomes (for $\KE=\mathbb{CP}^2$) the IIA reduction of $M^{3,2}$ \cite{witten-M32,castellani-dauria-fre-M32}, as worked out in \cite[(2.10)]{petrini-zaffaroni}.

\subsection{Product base}\label{sec:pro-base}
In this class the metric of $M_4$ splits as
\begin{equation}
\hat{g}^{(4)} = \hat{g}_1 + \hat{g}_2 \ .
\end{equation}
Accordingly, $\hat{\jmath} = \hat{\jmath}_1 + \hat{\jmath}_2$
and $\hat{\mathfrak{R}} = \hat{\mathfrak{R}}_1 + \hat{\mathfrak{R}}_2$, with
\begin{equation}\label{probase}
\hat{\mathfrak{R}}_1 = \frac{\kappa_1}{Q_1(y)} \hat{\jmath}_1 \ , \qquad \hat{\mathfrak{R}}_2 = \frac{\kappa_2}{Q_2(y)} \hat{\jmath}_2 \ ,
\qquad \kappa_1, \kappa_2 = \pm 1,\, 0 \ .
\end{equation}
The dependence of the metric of $M_4$ on $y$ is given by
\begin{equation}
\hat{g}^{(4)}(y,x^i) = Q_1(y) g_{\Sigma_1} + Q_2(y) g_{\Sigma_2}   \ ,
\end{equation}
where $g_{\Sigma_1}$,  $g_{\Sigma_2}$ are the metrics of the two Riemann surfaces $\Sigma_1$, $\Sigma_2$ of
scalar curvature $R_1 = 2 \kappa_1$ and $R_2 = 2 \kappa_2$ respectively.

When combined with \eqref{ricci}, and the fact that $\partial_y \hat{\mathfrak{R}} = 0$ the condition \eqref{probase} fixes
\begin{align}
Q_1 &= \frac{\kappa_1}{6} (3 \ell^2 y^{-1} - F_0^2 y^3) + \nu_1  \ , \\
Q_2 &= \frac{\kappa_2}{6} (3 \ell^2 y^{-1} - F_0^2 y^3) + \nu_2  \ ,
\end{align}
where $\nu_1$, $\nu_2$ are constants. In combination with \eqref{self-ricci}, \eqref{probase} gives the ODE $2T = (\kappa_1 Q_2 + \kappa_2 Q_1)/(Q_1Q_2)$ which, given the expression \eqref{T} for $T$ and defining again $p(y) =  e^{4A} y^2$, becomes a Riccati:
\begin{equation}\label{eq:pro-riccati}
y^2 \frac{2 Q_1 Q_2}{\kappa_1 Q_2 + \kappa_2 Q_1} \frac{dp}{dy} =  F_0^2 p^2 + (\ell^2-F_0^2 y^4) p - \ell^2 y^4 \ .
\end{equation}
This is solved by:
\begin{equation}\label{prosol}
p = \ell^2 y^2 \frac{ 3 \ell^2 \mu - 9 \kappa_1 \kappa_2 \ell^2 y^2  - 6 (\kappa_1\nu_2 + \kappa_2 \nu_1) y^3 +  \kappa_1 \kappa_2 F_0^2 y^6}{9 \kappa_1 \kappa_2 \ell^4 + 18 \ell^2 (\kappa_1\nu_2 + \kappa_2 \nu_1)  y + (36 \nu_1\nu_2 - 3 \ell^2 F^2_0 \mu) y^2 +  3 \kappa_1 \kappa_2 \ell^2 F_0^2 y^4} \ ,
\end{equation}
where $\mu$ is a constant parameter. The $\ell \to 0$ limit is well-defined after shifting $\mu \to 12 \nu_1 \nu_2/(F^2_0 \ell^2) + \mu$.

\subsubsection{Regularity and boundary conditions}\label{sec:pro}
We now turn to the analysis of the geometry of the solutions,  in a manner similar to the one in the previous section.

The metric \eqref{ClassAM6} on the internal manifold takes the form:
\begin{subequations}\label{eq:pro}
\begin{equation}\label{eq:pro-met}
e^{-2A} ds^2_{M_6} = - \frac14 \frac{q'}{x q} dx^2 - \frac{q}{x q' - 4 q} D\psi^2 + \frac{\kappa_1 q'}{x u_1} ds^2_{\Sigma_1}+ \frac{\kappa_2 q'}{x u_2} ds^2_{\Sigma_2} \ ,
\end{equation}
where $q$, $u_1$, $u_2$ are polynomials. The warp factor is given by
\begin{equation}\label{eq:pro-A}
L^{-2} e^{2A} = \sqrt{\frac{x^2q'-4xq}{q'}} \ .
\end{equation}
The dilaton is given  by
\begin{equation}\label{eq:pro-phi}
	g_s^{-2} e^{2 \phi} = \frac{q'}{x u_1 u_2} \left( \frac{x^2q'-4xq}{q'} \right)^{3/2} \ .
\end{equation}
\end{subequations}
The above is valid for $\kappa_1, \, \kappa_2 \neq 0$. When one of the two Riemann surfaces is flat, a slight modification
is required and we will treat the case $\kappa_2 = 0$  separately.

Notice that
\begin{equation}\label{eq:3qu1u2}
	3q'-xq''= \frac x2 (u_1 +  u_2) \ ;
\end{equation}
so we see that for $u_1 = u_2$ we recover (\ref{eq:ke}).

Positivity of the metric and the dilaton now requires either
\begin{subequations}\label{eq:pro-pos}
\begin{equation}\label{eq:pro-pos1}
	q<0 \ ,\qquad x q'>0 \ ,  \qquad \kappa_a u_a >0 \ , \qquad  u_1 u_2 > 0 \ ,
\end{equation}
or
\begin{equation}\label{eq:pro-pos2}
	q>0 \ ,\qquad x q'<0 \ ,  \qquad \kappa_a u_a <0 \ , \qquad  u_1 u_2 < 0 \ ,
\end{equation}
\end{subequations}
$a = 1,2$ (no summation over repeated indices). (\ref{eq:pro-pos1}) generalizes (\ref{eq:pos}), while (\ref{eq:pro-pos2}) is a new possibility.

Given the similarity between (\ref{eq:pro}) and (\ref{eq:ke}), most of the analysis leading to Table \ref{tab:bc} is the same. There is the additional possibility of the occurrence of a double zero. Moreover, the case of an inflection point now ramifies into three different branches. See Table \ref{tab:dz}.

\begin{itemize}
\item[]\textbf{Double zero: orbifold singularity.}
This did not occur in section \ref{sec:ke}, because a multiple zero was always a triple zero. This is no longer the case in the present section: a double zero which is not also triple can occur, and we must analyze it separately. A crucial fact is that $\Delta(q)\propto \mathrm{res}(q,u_1)\mathrm{res}(q,u_2)$. Thus when a double zero occurs either $u_1$ or $u_2$ has a zero. Choosing the latter, one finds the local expression for the metric
\begin{equation}
	\frac{1}{L^2 |x_0|} ds^2_{M_6} \sim \frac1{2x_0}\left[\frac{dx^2}{x_0-x}+ (x_0-x)\left(D \psi^2 - \frac{2\kappa_1 q''_0}{u_{10}} ds^2_{\Sigma_1}\right)\right] + \frac{\kappa_2 q''_0}{x_0 u'_{20}}ds^2_{\Sigma_2} \ ,
\end{equation}
while the warp factor and dilaton are constant. Here  $u_{10}\equiv u_1(x_0)$ and $u'_{20}\equiv u'_2(x_0)$. From \eqref{eq:3qu1u2} we see that $-2q''_0/u'_{10}=1$. Positivity of the metric requires that $x_0(x_0-x)>0$, and $\kappa_1=1$, which selects $\Sigma_1=S^2$.  With the choice $r=\sqrt{x_0-x}$, the parenthesis  becomes proportional to $dr^2 + \frac{1}{4} r^2 (D \psi^2 + ds^2_{S^2})$. If the $S^1$ periodicity is $\Delta \psi= 2\pi$, this is the local metric for an $\mathbb{R}^4/\mathbb{Z}_2$ singularity; if $\Delta \psi = 4 \pi$, it is $\mathbb{R}^4$, and we have a regular point.

\item[] \textbf{Inflection point at the origin: O4-, O6- or O8-plane.} In section \ref{sec:ke}, at an inflection point at the origin the denominator of the coefficient of $ds^2_\mathrm{\KE}$ has a double zero, canceling the double zero of the numerator, so that the full coefficient remains constant. In (\ref{eq:pro-met}), however, the functions $u_a$ in the denominator are independent of $q$. If neither of $u_a$ vanishes, the local metric and dilaton read
\begin{equation}
\begin{split}
	\frac{1}{L^2}\sqrt{-\frac{q_3}{8q_0}}ds^2_{10}&\sim \frac1{\sqrt x}\left(ds^2_{\mathrm{AdS}_4} + \frac{1}{4}D \psi^2\right)
	-\frac{q_3}{8q_0} \sqrt x \left(dx^2 - \frac{4q_0 \kappa_1}{u_{10}} ds^2_{\Sigma_1}-\frac{4q_0 \kappa_2}{u_{20}} ds^2_{\Sigma_2}  \right) \ , \\
	g_s^{-2} e^{2\phi}&\sim \frac{-4q_0}{u_{10}u_{20}} \sqrt{-\frac{8q_0}{x q_3}} \ ,
\end{split}
\end{equation}
According to the discussion underneath \eqref{eq:O4}, this locus describes an O4-plane smeared over $\Sigma_1\times \Sigma_2$.

If only one of the $u_a$, say $u_1$, vanishes,\footnote{Equation (\ref{eq:3qu1u2}) appears to imply that  at an inflection point either the $u_a$ both vanish or both do not vanish. However, we will see in section \ref{sub:k20} that for $\kappa_2=0$ this formula is modified, so that only $u_1$ needs to vanish.}
 we have
\begin{equation}
	\begin{split}
		L^{-2}\sqrt{-\frac{q_3}{8q_0}}ds^2_{10}&\sim \frac1{\sqrt x}\left(ds^2_{\mathrm{AdS}_4} + \frac{1}{4}D \psi^2 + \frac{q_3\kappa_1}{2 u'_{10}} ds^2_{\Sigma_1}\right)
		-\frac{q_3}{8q_0} \sqrt x \left(dx^2 -\frac{4q_0 \kappa_2}{u_{20}} ds^2_{\Sigma_2}  \right) \ , \\
		g_s^{-2}e^{2\phi}&\sim \frac{q_3}{2u'_{10}u_{20}} \left( -\frac{8q_0}{x q_3} \right)^{3/2} \ .
	\end{split}
\end{equation}
This locus corresponds to an O6-plane singularity. Adapting our discussion below (\ref{eq:O4}), we conclude that it is localized if $\Sigma_2=S^2$ ($\kappa_2=1$), while it is smeared over $\Sigma_2$ if $\kappa_2 = -1$.

Finally, if both $u_a$ vanish at the inflection point, there is an O8-plane singularity as in \eqref{eq:o8}, with the $\KE$ base replaced by $\Sigma_1\times \Sigma_2$.

\item[] \textbf{Quartic maximum at the origin: O4/O8-plane.} When $q \sim q_0 + \frac{1}{4!} q_4 x^4$, then both $u_a$ have a single zero. We have
\begin{equation}
	\begin{split}
		L^{-2} \sqrt{-\frac{q_4}{24 q_0}}ds^2_{10}&\sim \frac1x \left(ds^2_{\mathrm{AdS}_4} + \frac{1}{4}D \psi^2 \right)
		+\frac{q_4}{6}\left( \frac{\kappa_1}{u'_{10}}ds^2_{\Sigma_1} + \frac{\kappa_2}{u'_{20}}ds^2_{\Sigma_2} \right)
		-\frac{q_4}{24 q_0} x dx^2 \ , \\
		g^{-2}_s e^{2\phi} &\sim \frac{8\sqrt{6}}{u'_{10} u'_{20}}\sqrt{-\frac{q_0^3}{q_4}} x^{-3}\ .
	\end{split}
\end{equation}
which is the appropriate behavior for an O4/O8-plane singularity. This case occurs only if $\kappa_a$ have opposite signs, so the O4-plane is smeared over one of the $\Sigma_a$.

\begin{table}
\begin{center}
\begin{tabular}{cccccccc}
$x_0$ &$q(x_0)$ & $q'(x_0)$ & $q''(x_0)$ & $q'''(x_0)$ & $u_1(x_0)$ & $u_2(x_0)$ & interpretation \\\hline\hline
  & 0 & 0 &   &   & 0 &   &   $\mathbb{R}^4/\mathbb{Z}_2$ \\\hline
0 &   & 0 & 0 &   &   &   &  O4 \\\hline
0 &   & 0 & 0 &   & 0 &   &  O6 \\\hline
0 &   & 0 & 0 &   & 0 & 0 &  O8 \\\hline
0 &   & 0 & 0 & 0 & 0 & 0 &  O8/O4\\\hline
\end{tabular}
\end{center}
\caption{Additional singularities that occur in the product base case. The O8-plane one, which already appeared in Table \ref{tab:bc}, is repeated for comparison.}
\label{tab:dz}
\end{table}
\end{itemize}

\subsubsection{Generic case} 
\label{sub:genpro}
Here we define new parameters
\begin{equation}\label{eq:gammaa-beta}
\gamma_a=\frac{1}{\kappa_a}\frac{1}{\ell^2}\sqrt{\frac{\sqrt 3 \ell}{F_0}}\nu_{a} \ , \qquad
\beta = \frac{1}{\kappa_1\kappa_2}\frac{F_0}{2\sqrt{3}\ell} \mu \ ,
\end{equation}
$a = 1,2$, and a new coordinate $x= \sqrt{\frac {F_0}{\sqrt 3 \ell}} y $. The solution is then cast in the form (\ref{eq:pro}), with
\begin{equation}
   \begin{split}
   	 q&= x^6 + 3(2\gamma_1 \gamma_2 - \beta) x^4 + 4 (\gamma_1+\gamma_2) x^3 + 3 x^2 - \beta \ , \\
	  u_1&=12(1+2\gamma_1 x-x^4)   \, ,\qquad u_2=12(1+2\gamma_2 x-x^4) \,
   \end{split}
\end{equation}
and the constants $L$, $g_s$ determined by (\ref{eq:Lgsgen}).

We again note that the Riccati ODE (\ref{eq:pro-riccati}) implies an ODE directly on $q$, $u_1$, $u_2$:
\begin{equation}\label{eq:pro-qode}
	\frac x{12}(3q'-xq'')^2 = -\left(\frac1{u_1}+\frac1{u_2}\right)\left[(xq'-4q)(1+3x^4)+4q\right]\,,
\end{equation}
similar to (\ref{eq:qode})

In this case we will not give an exhaustive discussion as in section \ref{sub:gen} for the K\"{a}hler--Einstein base. There are many different possibilities, and a full discussion would be rather tedious. Let us instead give here a summary of the main features of the parameter space.

As in the K\"{a}hler--Einstein base case, the interpretation of the solution depends on the properties of the polynomial $q$, and less importantly on $u_1$, $u_2$. The most important features are the presence of extrema, and the presence of zeros. These can be decided again by looking at the discriminants of $q$ and $q'$. Unlike in section \ref{sec:ke}, these are unrelated, and vanish on different loci of parameter space. It is also useful to notice that $2^{26} 3^{10}\Delta(q)= \beta \, \mathrm{res}(q',u_1)\mathrm{res}(q',u_2)\propto \Delta(q)$; thus, the sign of $u_1$, $u_2$ at an extremum of $q$ (which has to do with the sign of $\kappa_1$, $\kappa_2$) changes on the discriminant of $q$. The expressions of the resultants are
\begin{align}\label{eq:res12}
		\mathrm{res}(q',u_1)&= - 2^{18}3^9 \left( \beta^4 - 6 \gamma_1 \gamma_2 \beta^3+ (12 \gamma_1^2 \gamma_2^2-2) \beta^2 -2(2 \gamma_1^2 + \gamma_1 \gamma_2 +2 \gamma_2^2+ 4 \gamma_1^3 \gamma_2^3) \beta +1 -\gamma_1^4 +2 \gamma_1^3 \gamma_2\right) \, ,
		\, \nonumber \\
		\mathrm{res}(q',u_2)&= \mathrm{res}(q',u_1) + 2^{18}3^9 (\gamma_1 + \gamma_2) (\gamma_1 - \gamma_2)^3\,.	
\end{align}

Depending on the signs of $\kappa_a$, there are three possibilities to consider: (i)  For $\kappa_a =-1$ there are solutions
with a regular endpoint and an endpoint with an O4-plane singularity. (ii)  For $\kappa_a =+1$ there are regular solutions and ones with orbifold singularities which were found numerically in \cite{tz}. There are also solutions which
for $\Delta\psi = 4\pi$ have $\mathbb{CP}^3$ topology and were found numerically in \cite{ajtz}. There are also solutions with one or two O4-plane singularities.
(iii) For $\kappa_1=+1$, $\kappa_2=-1$ (or viceversa) there are solutions with one or two O4-plane singularities, as well as
solutions with an orbifold singularity.

The flux potentials are again defined by \eqref{twistflux}, now with $\hat{\jmath} = Q_1 j_1 + Q_2 j_2$. Using $\rho = \rho_1 + \rho_2$, 
with $d_4 \rho_a = - \kappa_a j_a$ (no summation),  we find
\begin{equation}
C_1 = f_2 D \psi + \frac12(\kappa_1 \nu_1 - \kappa_2 \nu_2)(\rho_1 - \rho_2)  \, , \quad
C_3 =  D\psi \wedge (g_1 j_1 + g_2 j_2) 
\, , \quad
C_5 = f_6 D\psi \wedge j_1 \wedge j_2 \, ,
\end{equation} 
 where
\begin{subequations}\label{eq:C35xqpro}
\begin{align}
f_2 &= 2\ell\sqrt{\frac{F_0}{\sqrt 3 \ell}}\left(\frac1x\frac q{4q-xq'}-\frac{3q'-x q''}{48x^2}\right) \, , \\
g_a &=\frac{\ell^3}{8 \sqrt3 F_0}\int x u_a dx \, , \\
f_6 &= \frac{\ell}{72}\sqrt{\frac{\sqrt{3}\ell}{F_0}}\left[q'-\frac x{768}(3u_1^2-10u_1 u_2 + 3u_2^2)+x\left(3+x^4+\frac{u_1+u_2}{16}\right)^2\right] \, .
\end{align}
\end{subequations}
The integration constant in the indefinite integral for the $g_a$ is chosen such that
\begin{equation}
g_a(x=0)= -\frac{\ell^3}{8 \sqrt3 F_0} \frac{\beta}3 \ .
\end{equation}

\subsubsection{$\ell=0$} 
\label{sub:l0pro}

In this case we introduce
\begin{equation}
	\s= - \frac{3 \mu}{y_0^2}\ ,\qquad t= \frac{\kappa_2 \nu_1}{\kappa_1 \nu_2}\ ,\qquad x= \frac y{y_0}
          \ , \qquad y_0 \equiv \left(\frac{6 \nu_2}{\kappa_2 F_0^2}\right)^{1/3} \ .
\end{equation}
The solution is now (\ref{eq:pro}) with
\begin{equation}\label{eq:ql0pro}
	q=x^6+\frac{\s}{2} x^4+2(1+t)x^3-\frac{t}{2} \ ,\qquad u_1 =12 x(1-x^3) \ ,\qquad u_2 =12 x(t-x^3)\ ,
\end{equation}
and the constants $L^2= |y_0|$, $g^2_s = 72/(|y_0| F_0^2)$.

For $t=1$, we recover the solutions in section \ref{sub:l0}, with $\KE=\mathbb{CP}^1\times \mathbb{CP}^1$. For other values of $t$, the analysis is similar.

Define
\begin{equation}\label{eq:s1+-}
	\s_1 \equiv -9\cdot 4^{-1/3} |1+t|^{2/3} \ ,\qquad
	\s_+ \equiv -3(2+t) \ ,\qquad \s_- \equiv -3 t^{-1/3}(1+2t) \ .
\end{equation}
The discriminant $\Delta(q)= 2t^3(\s^3-\s_+^3)(\s^3-\s_-^3)$, while $\Delta\left(q'/x^2\right)=-192 (\s^3-\s_1^3)$. In terms of these, the following solutions are possible:
\begin{itemize}
   \item $\kappa_a=+1$:
	With a single O4-plane singularity for $t<0$, $\s<\s_-$, and also for $t>0$ and any $\s$. With an O8-plane singularity for $t>0$; for $\s>\s_-$ it also has an O4-plane singularity, for $\s=\s_-$ it has an orbifold singularity, for $\s<\s_-$ no other singularity.
	\item $\kappa_a=-1$:
	With a single \Of for $\s<\s_+$ and any $t$.
	
	\item $\kappa_1=+1$, $\kappa_2=-1$:
	With two O4-plane singularities for $\s_+<\s<\s_-$; with an O4-plane singularity and an orbifold singularity for $\s=\s_+$; with a single O4-plane singularity for $t>0$, $\s_-<\s<\s_+$, or for $t<0$, $\s<\s_+$.
	
	With an O8-plane singularity for $-1<t<0$; for $\s>\s_-$ it also has an O4-plane singularity, for $\s=\s_-$ it has an orbifold singularity, for $\s<\s_-$ no other singularity.

	With an O8/O4-plane singularity for $t=-1$; for $\s<0$ it also has an O4-plane singularity, for $\s=-1$ an orbifold singularity, for $\s<-1$ no other singularity.
\end{itemize}

\subsubsection{$\kappa_2=0$} 
\label{sub:k20}

In this case we need to adjust the form of the solution as presented in \ref{sec:pro}, by replacing the factor $\kappa_2$ in front of the line element of $\Sigma_2$ by a constant $m$, to be specified below.
The functions that determine the solution read:
\begin{equation}\label{eq:qk20}
	q= 3 x^4 - 4 x^3 +n \ ,\qquad u_1=-24 x \ , \qquad u_2 = 3\tilde\ell^2(1-x) + n x - x^4 \
\end{equation}
where the coordinate $x = y/L^2$, with $L^2$ related to the constant parameters  appearing in \eqref{prosol} as:
\begin{equation}
L^2 = \frac{6 \ell^2 \kappa_1 \nu_2}{F_0^2 \ell^2 \mu - 12 \nu_1 \nu_2}\,.
\end{equation}
For $n$, $\tilde\ell$ and $m$   we have the following relations:
\begin{equation}
\tilde{\ell} = \frac{\ell}{L^4 F_0}
\ , \qquad
n = \frac{6 \nu_1}{\kappa_1 L^6 F_0^2} + 3\ell^2
\ , \qquad
m = -\frac{4 F_0^2 L^6}{\nu_2} \ .
\end{equation}
Finally, $g_s^2 = 2/(\sqrt{3} F_0^2 L^8)$.

In this case:
\begin{itemize}
	\item $\kappa_1=+1$: there is a solution with an O6-plane smeared along the $T^2$, for $\tilde \ell \neq 0$. This becomes an O8-plane for $\tilde \ell=0$. For $n>1$ it also has an O4-plane singularity; for $n=1$ it has an orbifold singularity; for $n<1$, $n\neq 0$ no other singularity.
	 \item $\kappa_1=-1$: there is a solution with a single O4-plane singularity for $n<1$.
\end{itemize}

\subsubsection{$F_0=0$} 
\label{ssub:prof0}

Similarly to section \ref{sub:f00}, we define the coordinate $x=\frac{\nu_2}{\ell^2} y$; this differs from the $x$ defined in section \ref{sub:gen} by a factor $\gamma_2$, recalling the definitions in (\ref{eq:gammaa-beta}). Let us define $t_i \equiv \gamma_i \sqrt \beta$. Taking the limit $\beta\to 0$ with $t_i$ kept constant yields now the solution (\ref{eq:pro}) with
\begin{equation}\label{eq:qLgsf00pro}
	q=\frac{t_1}{t_2} x^4+\frac23 \left(1+ \frac{t_1}{t_2}\right) x^3+\frac12 x^2-\frac{t_2^2}6 \, ,\qquad
	u_1= 2+4x \, ,\qquad u_2 =2+4 \frac{t_1}{t_2}x \,.
\end{equation}
Moreover $L^2=\ell^2 \kappa_1 t_2/(\nu_1 t_1)$, $g_s=2 L^3/\ell$.

\begin{figure}[ht]
	\centering
		\includegraphics[height=3in]{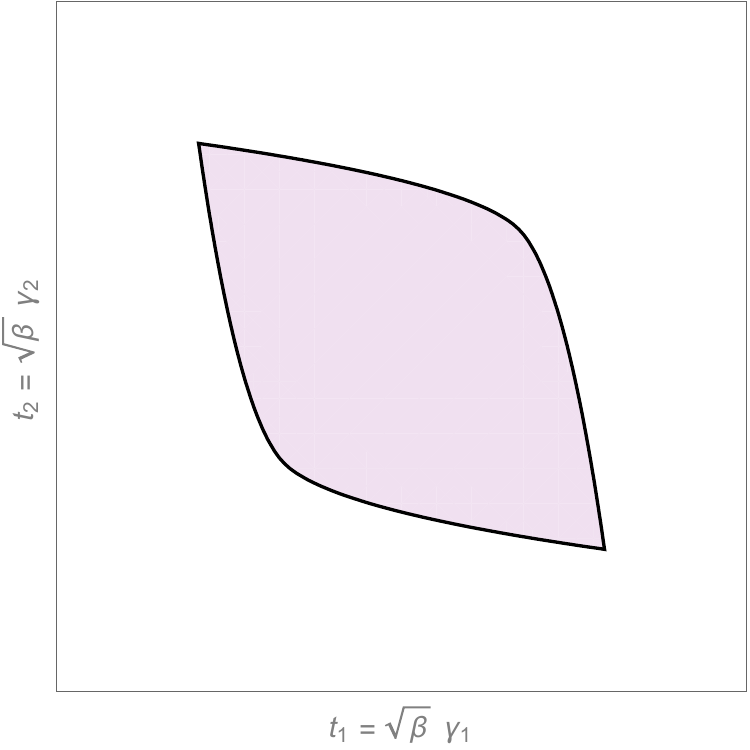}
	\caption{The parameter space of regular solutions with $F_0 = 0$ for the product base case.}
	\label{fig:product-massless}
\end{figure}

Regular solutions exist for $\kappa_a=+1$. The parameter space is obtained by considering the equations for the resultants in (\ref{eq:res12}), after taking $\beta\to0$ with $t_i$ kept constant; this gives two cubics, and the allowed parameter space is enclosed by them. The result is shown in figure \ref{fig:product-massless}. (Points related by inversion $t_1\leftrightarrow t_2$ correspond to the same solution.) These correspond to the solutions found in \cite[Sec.~4.5]{gauntlett-martelli-sparks-waldram-Apqr} and \cite{chen-lu-pope-vazquezporitz}, more or less in the coordinates of the first reference. They were studied in more detail in \cite[Sec.~3]{tz} where they were referred to as $A^{p,q,r}$ solutions. For $t_1=t_2$ we recover the case of section \ref{sub:f00} for $\KE=\mathbb{CP}^1\times \mathbb{CP}^1$. At the boundary of the region in figure \ref{fig:product-massless}, $M_6$ no longer has the topology of an $S^2$-fibration over $\mathbb{CP}^1\times \mathbb{CP}^1$. For example, the solution at $t_1=-t_2 = \frac{\sqrt3}{2\sqrt2}$ is the Fubini--Study solution on $\mathbb{CP}^3$. We refer to \cite[Sec.~3]{tz} for a detailed discussion.


\subsection{Summary} 
\label{sub:sum}

We will now summarize the physically meaningful solutions we have found. This means we will exclude the solutions with smeared orientifolds, which are of dubious physical significance. The only fully-localized orientifolds we could find are O8-planes. We will first discuss solutions without any O8-planes, and then solutions with O8-planes.

\begin{figure}[ht]
	\centering
		\includegraphics[height=7cm]{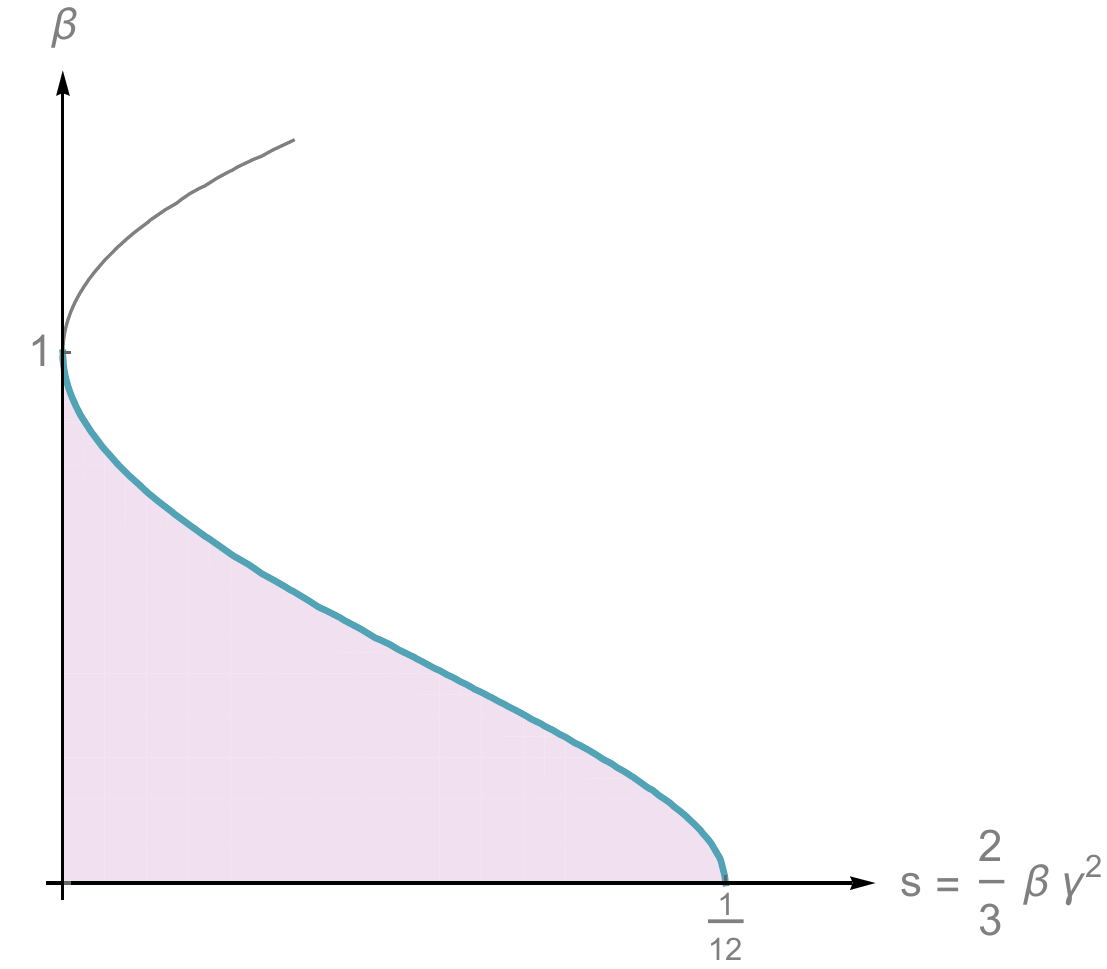}
	\caption{The allowed parameter space for regular solutions in the generic case with $\KE$ base is the interior of the purple region.}
	\label{fig:KE-summary}
\end{figure}

\textbf{Solutions with $\KE$ base and without orientifolds.} The parameter space is shown in figure \ref{fig:KE-summary}, which is a different version of figure \ref{fig:KE-plots}. In all this subsection the $\psi$ periodicity is $\Delta \psi= 2\pi$, unless otherwise stated.
\begin{itemize}
	\item The purple region corresponds to fully regular solutions. It is given by $\beta>0$ and
	\begin{equation}\label{eq:beta-s}
		\beta^6 - (2+9 s)\beta^4 + (1-15s+27s^2) \beta^2 +\frac94 s^2(1-12s)>0 \,.
	\end{equation}
	Recall that $s$ is related to $\beta$ and $\gamma$ by (\ref{eq:s-bg}).
	
	 Topologically, $M_6$ is an $S^2$-fibration over $\KE$.
	
	These solutions realize the numerical solutions found in \cite{petrini-zaffaroni} and later generalized by \cite{lust-tsimpis-singlet-2,tz}.
	
	\item The boundary $\beta=0$ of the purple region gives $F_0=0$; these are IIA reductions of the $Y^{p,k}$ Sasaki--Einstein solutions in \cite{gauntlett-martelli-sparks-waldram-SE7}. Indeed from (\ref{eq:beta-s}) we see that $0<s<\frac1{12}$, as derived there. The limit $s\to0$  can be taken only after rescaling $s\to s/\nu^2$, in which case it leads to the IIA reduction \cite[(2.10)]{petrini-zaffaroni} of the $M^{3,2}$ Sasaki--Einstein \cite{witten-M32,castellani-dauria-fre-M32}, generalized by replacing $\mathbb{CP}^2$ with an arbitrary $\KE$.
	
	\item The green boundary (obtained by replacing the inequality with equality in (\ref{eq:beta-s})) corresponds to solutions with a conical Calabi--Yau singularity. For $\KE=\mathbb{CP}^2$, this is an orbifold singularity $\mathbb{C}^3/\mathbb{Z}_3$.

	\item The point at $\beta=1$, $s=0$ has two conical Calabi--Yau singularities. For $\KE=\mathbb{CP}^2$, these are both orbifold singularities $\mathbb{C}^3/\mathbb{Z}_3$. In this case, however, it is also possible to take $\Delta \psi=6\pi$, in which case the space becomes fully regular; this is the Guarino--Jafferis--Varela solution \cite{guarino-jafferis-varela}, whose generalization for arbitrary $\KE$ was considered in \cite{fluder-sparks}.
	
\end{itemize}

\begin{figure}[ht]
	\centering
		\includegraphics[height=3in]{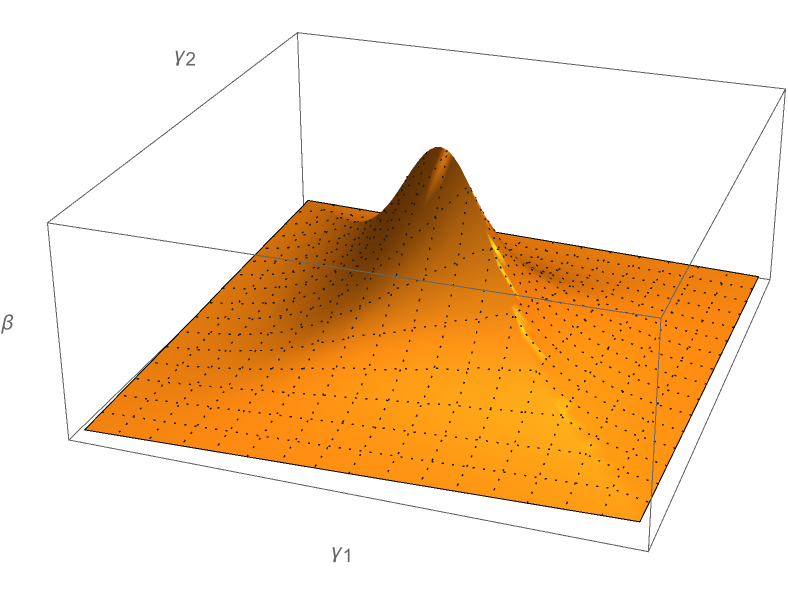}
	\caption{The allowed moduli space for regular solutions in the generic case with product base is between the plotted surface and the plane $\beta=0$.}
	\label{fig:product-summary}
\end{figure}

\textbf{Solutions with product base and without orientifolds.} There are regular solutions with $\kappa_a=+1$. The parameter space is shown in figure \ref{fig:product-summary}, which is defined by certain branches of $\mathrm{res}(q',u_a)=0$ (see (\ref{eq:res12})). This time we use the original parameters $\gamma_1$, $\gamma_2$, $\beta$.
These solutions were discussed in \cite[Sec.~5]{tz} in detail, although they were only known numerically in that paper; thus we will be brief.

\begin{itemize}
	\item The region in figure \ref{fig:product-summary} between the plotted surface and $\beta=0$ corresponds to regular solutions with the topology of an $S^2$-bundle over $\mathbb{CP}^1\times \mathbb{CP}^1$.
	\item The limit $\beta \to 0$, with $t_a = \gamma_a \sqrt \beta$ kept constant, reproduces the $A^{p,q,r}$ massless solutions whose parameter space was shown in figure \ref{fig:product-massless}.
	\item The boundary of figure \ref{fig:product-summary} corresponds to solutions with a $\mathbb{Z}_2$ orbifold singularity.
	\item The locus $\{\gamma_1=\gamma_2\}$ is a particular case of the solutions with $\KE$ base, where $\KE= \mathbb{CP}^1\times \mathbb{CP}^1$. At the boundary, there is a conifold$/\mathbb{Z}_2$ singularity.
	\item The intersection of the boundary with the locus $\{\gamma_1 = - \gamma_2\}$ (visible as the ridge in figure \ref{fig:product-massless}) corresponds to solutions with topology $\mathbb{CP}^3/\mathbb{Z}_2$. In this case one also has the option of taking $\Delta \psi= 4\pi$, thus making the topology directly $\mathbb{CP}^3$. These are the solutions studied in \cite{ajtz}.
	\item The point $\beta=1$, $\gamma_a=0$ is a variant of the solution in \cite{guarino-jafferis-varela}, with $\mathbb{CP}^2$ replaced by $\mathbb{CP}^1\times \mathbb{CP}^1$.
\end{itemize}

\textbf{Solutions with O8-planes.} There are many solutions which are regular except for a single O8-plane singularity. They occur for $\ell=0$, both for $\KE$ and for product base. Here is a list of possibilities:

\begin{itemize}
	\item $\KE$ base with  $\kappa=+1$, and $\s>-9$ in (\ref{eq:qLgsl0}).
	\item Product base with $\kappa_1=+1$, $\kappa_2 = \pm 1$, and $\s>\s_-$; see (\ref{eq:ql0pro}), (\ref{eq:s1+-}).
	\item Product base with $\kappa_1=+1$, $\kappa_2=0$, and $n<1$; see (\ref{eq:qk20}).
\end{itemize}


\section*{Acknowledgements}
AP is supported by the Knut and Alice Wallenberg Foundation under grant Dnr KAW 2015.0083. DP and AT are supported in part by INFN.

\appendix

\bibliography{at}
\bibliographystyle{utphys}

\end{document}